\PassOptionsToPackage{pdfpagelabels=false}{hyperref} 
\documentclass[a4paper,fleqn,usenatbib]{mnras}

\usepackage[T1]{fontenc}
\usepackage{ae,aecompl}

\usepackage{graphicx}	% Including figure files
\usepackage{amsmath}	% Advanced maths commands
\usepackage{amssymb}	% Extra maths symbols
\usepackage{multicol}   % Multiple Columns
\usepackage{longtable} % Table covering multiple pages
\usepackage{caption} % Caption setup
\usepackage{lscape}
\title[Uncovering the Orbit of the Hercules Dwarf]{Uncovering the Orbit of the Hercules Dwarf Galaxy}

\author[A. L. Gregory et al.]{Alexandra L. Gregory$^{1}$\thanks{E-mail: a.gregory@surrey.ac.uk},
Michelle L. M. Collins$^{1}$,
Denis Erkal$^{1}$,
Erik Tollerud$^{2}$,
\newauthor Maxime Delorme$^{1}$,
Lewis Hill$^{1}$,
David J. Sand$^{3}$,
Jay Strader$^{4}$
Beth Willman$^{5}$,
\\
$^{1}$Department of Physics, University of Surrey, Guildford, GU2 7XH, Surrey, UK\\
$^{2}$Space Telescope Science Institute, 3700 San Martin Dr, Baltimore, MD 21218, USA\\
$^{3}$Department of Astronomy/Steward Observatory, 933 North Cherry Avenue, Rm. N204, Tucson, AZ 85721-0065, USA\\
$^{4}$Department of Physics and Astronomy, Michigan State University, East Lansing, MI 48824, USA\\
$^{5}$National Optical-Infrared Astronomy Research Laboratory, 950 North Cherry Avenue, Tucson, AZ 85719, USA
}

\date{Accepted XXX. Received YYY; in original form ZZZ}

\pubyear{2019}

\begin{document}
\label{firstpage}
\pagerange{\pageref{firstpage}--\pageref{lastpage}}
\maketitle

\begin{abstract}
\noindent We present new chemo--kinematics of the Hercules dwarf galaxy based on Keck II-- DEIMOS spectroscopy. Our 21 confirmed members have a systemic velocity of $v_{\mathrm{Herc}}=46.4\pm1.3$ kms$^{-1}$ and a velocity dispersion $\sigma_{v,\mathrm{Herc}}=4.4^{+1.4}_{-1.2}$ kms$^{-1}$. From the strength of the Ca II triplet, we obtain a metallicity of [Fe/H]= $-2.48\pm0.19$ dex and dispersion of $\sigma_{\rm{[Fe/H]}}= 0.63^{+0.18}_{-0.13}$ dex. This makes Hercules a particularly metal--poor galaxy, placing it slightly below the standard mass--metallicity relation. Previous photometric and spectroscopic evidence suggests that Hercules is tidally disrupting and may be on a highly radial orbit. From our identified members, we measure no significant velocity gradient. By cross--matching with the second \textit{Gaia} data release, we determine an uncertainty--weighted mean proper motion of $\mu_{\alpha}^*=\mu_{\alpha}\cos(\delta)=-0.153\pm{0.074}$ mas yr$^{-1}$, $\mu_{\delta}=-0.397\pm0.063$ mas yr$^{-1}$. This proper motion is slightly misaligned with the elongation of Hercules, in contrast to models which suggest that any tidal debris should be well aligned with the orbital path. Future observations may resolve this tension.

\end{abstract}


%%%%%%%%%%%%%%%%%%%%%%%%%%%%%%%%%%%%%%%%%%%%%%%%%%

%%%%%%%%%%%%%%%%% BODY OF PAPER %%%%%%%%%%%%%%%%%%

\section{Introduction} \label{sec:intro}

The recently discovered class of ultra--faint dwarf galaxies (UFDs) denote the low--mass regime of galaxy formation. Initially identified as over--densities in wide--field surveys such as the Sloan Digital Sky Survey (SDSS, \citealt{york00}), UFD detections have more than doubled the number of known Milky Way satellites (e.g. \citealt{bechtol15,drlicawagner15,koposov15}, among others). They appear to have unusually low surface brightnesses and luminosities, which, combined with high dynamical masses, suggests that they may represent the most dark matter dominated systems in the Universe \citep{simon07}. However, the lack of bright stars renders observations, and in particular follow--up spectroscopy, difficult with current facilities.

\begin{figure*}
	\centering
	\includegraphics[width=\linewidth] {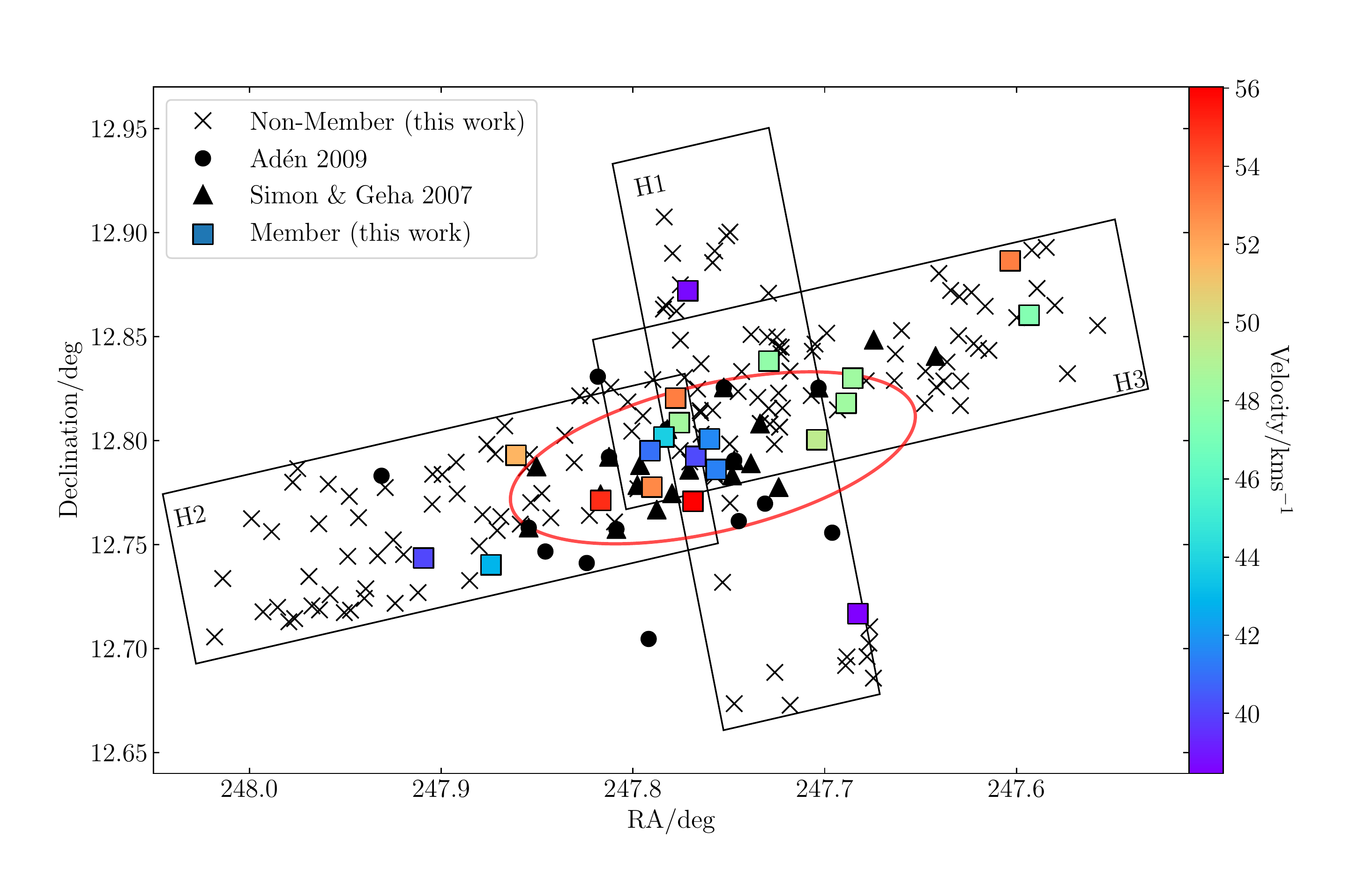}
	\caption{RA--Dec. map of Hercules. The targets observed with Keck/ DEIMOS in this work are shown as black crosses, with those identified as members highlighted as squares colour coded by velocity. The approximate positions of our DEIMOS masks are outlined, and the red ellipse marks the nominal half--light radius of Hercules. The positions of the identified members (with radial velocity measurements available) observed by \citet{simon07} (triangles) and \citet{aden09a} (circles) are shown. The systemic velocity of Hercules is 46.4$\pm$1.3kms$^{-1}$.}
	\label{fig:obsmap}
\end{figure*}

The shape of the mass profiles of these galaxies may play a role in their ability to survive tidal interactions. A key prediction of pure dark matter $\Lambda$CDM simulations is that dwarf galaxies should play host to a steep central cusp in the density profile \citep{dubinski91,navarro96}. The stars in such a galaxy would be deeply embedded in the halo, enabling the galaxy to withstand tidal disruption and stripping \citep{penarrubia08}. However, observations suggest that shallow cores may be far more prevalent among Milky Way dwarf spheroidals \citep{moore94,walker11,amorisco13}, and also among ultra--faint dwarfs \citep{amorisco17,contenta18}, making them much more susceptible to tidal disruption. Recent work by \citet{read16} demonstrated that an extended period of bursty star formation is capable of reducing a cusp to a core through dark matter heating (see also \citealt{read05,pontzen12,pontzen14,teyssier13,brooks13}). It is therefore likely that a large proportion of the Milky Way satellite population is cored, and, by extension, a significant number of satellites may be expected to be undergoing some form of tidal disturbance. Deeper studies into the number and nature of disrupting satellites (e.g. \citealt{mutlupakdil19}) may provide important clues as to the shapes of their density profiles, and the role of baryons in galaxy evolution.

To date, a number of ultra--faint dwarf galaxies have displayed evidence for tidal interactions (e.g. \citealt{martin08,munoz10,simon07}-- hereafter SG07; \citealt{sand12,collins17}, among others). One of the strongest candidates for a tidally disrupting UFD around the Milky Way is the Hercules dwarf galaxy. First identified by \citet{belokurov07} as a stellar overdensity in imaging data from SDSS, and located 132 kpc from the Sun \citep{musella12}, Hercules appears to be rapidly receding from the Milky Way ($v_{\rm{GSR}}\sim145$ kms $^{-1}$; SG07, \citealt{aden09a}-- hereafter A09). The same spectroscopic studies have shown Hercules to have a small velocity dispersion of $\sigma_v<5$ kms$^{-1}$, corresponding to a mass within 300 pc ($\approx$r$_{\rm{half}}$) of $M$(<300)$\sim2\times10^6M_{\odot}$ \citep{aden09b}.

\citet{sand09} assess the star formation history of Hercules using CMD fitting, and find an old stellar population with negligible star formation in the last 12 Gyr. This view was upheld by \citet{brown14}, supporting the hypothesis that star formation in the smallest halos is suppressed by reionization. Indeed, a study of HST imaging by \citet{weisz14} suggests that Hercules may be a `fossil' galaxy quenched by reionization. However, this scenario is complicated by the possible tidal disruption of Hercules. Like other ultra--faint satellites, Hercules appears to be metal poor with a large metallicity dispersion (SG07; \citealt{kirby08a}; A09). \citet{koch08} suggested that Hercules experienced inhomogeneous chemical enrichment, primarily due to the contribution of Type II supernovae.

There is photometric and spectroscopic evidence to suggest that Hercules is undergoing tidal disruption by the Milky Way. \citet{coleman07} find that Hercules has a 3:1 axis ratio \citep{martin08,sand09}, making it the most elongated Milky Way dSph other than the disrupting Sagittarius dwarf \citep{ibata94}. Deep imaging by \citet{roderick15} highlights at least nine significant over--densities as far as 2 kpc from the centre, mostly distributed along the major axis, though with some perpendicular to this. They find that the same number of stars are located in this substructure as in the main body of Hercules, suggesting a high rate of mass loss. \citet{deason12} spectroscopically confirm several blue horizontal branch (BHB) members at large distances, which are likely to have been tidally stripped from the galaxy. Similarly, \citet{garling18} detect three RR Lyrae members outside the nominal tidal radius. Given that there are nine RR Lyrae members within the tidal radius, this implies that a large proportion of stellar material has been stripped from the main body of Hercules. However, combining kinematics with proper motions from \textit{Gaia} DR2, \citet{fu19} were unable to confirm any members located in potential tidal substructures.

Considering the large heliocentric distance of Hercules--- one of the outermost known Milky Way satellites--- a highly eccentric orbit is required to bring it close enough to the Milky Way to induce tidal disruption. Work by \citet{fu19} suggests a 40$\%$ probability that Hercules has undergone tidal stripping, based on the principle that its orbit must bring it within 40kpc of the Milky Way (the distance at which the tidal radius is equal to 3$\times$ the half--light radius). It has been argued in multiple studies that if the elongation is the result of Hercules dissolving into a stellar stream, one would expect the orbit to be aligned with the major axis. \citet{martin10} measure a tentative velocity gradient of $\frac{dv_{r}}{d\chi}=10.2\pm6.0$ kms$^{-1}$ kpc$^{-1}$ along the major axis. From this, they constrain a pericentre radius of $R_{\mathrm{peri}}=6^{+9}_{-2}$ kpc, reached some 0.6 Gyr ago, consistent with a highly radial orbit and the possibility that Hercules is a stellar stream in formation.

However, analysis by \citet{blana15} suggests that the orbit described above is not feasible given the highly elongated appearance of the galaxy. The positions of the substructures identified in \citet{roderick15} are also suggestive of an orbital path which does not align exactly with the tidal arms. Others have proposed that the stream--in--formation is actually aligned with the minor axis of Hercules. \citet{kupper17} (hereafter K17) model a highly eccentric orbit with $\epsilon\approx0.95$, which causes the satellite to `explode' such that the tidal debris spreads out perpendicular to the orbital path. In this situation, the galaxy would experience tidal shocks at the pericentre of its orbit, which remove large portions of mass; one condition of this would be a particularly low central density. The dispersion of stars resulting from these tidal shocks mean that Hercules would be travelling in the direction of the apparent minor axis, forming a stream which is broader than it is long. The distribution of RR Lyrae stars detected in \citet{garling18} supports this hypothesis.

In this work, we re--examine the chemo--kinematics of Hercules using new spectroscopy from the DEIMOS instrument on the Keck II telescope, and combine this with proper motion data from \textit{Gaia} DR2 to investigate its potential orbit. The paper is structured as follows. In $\S$\ref{sec:obs} we detail the observations and data reduction process for our photometric and spectroscopic datasets, and outline the definition of Hercules membership. Our refined measurements of the kinematics and metallicity of Hercules is discussed in $\S$\ref{sec:chemodyn}. We explore the proper motion and orbital parameters of Hercules as measured in \textit{Gaia} DR2 in $\S$\ref{sec:gaia}. We conclude in $\S$\ref{sec:conc}. Key properties of Hercules as used in this work are outlined in Table \ref{tab:herc_properties}.

\begin{table}
	\centering
	\caption{Key Properties of Hercules. Properties in the lower panel are derived in this work. a) \citet{belokurov07}; b) \citet{musella12}; c) \citet{sand09}; d) \citet{aden09a}; e) \citet{martin08}. The orbital properties are derived in Sec. \protect\ref{sec:orbit}.}
	\begin{tabular}{|l|l|}
		\hline
		\textbf{RA$^{a}$} & 16 31 02\\[3pt]
		\textbf{Declination$^{a}$} & +12 47 30\\[3pt]
		\textbf{Heliocentric Distance$^{b}$} & 132$\pm$6 kpc\\[3pt]
		\textbf{Half--Light Radius$^{c}$} & 6.3$\pm$0.5 arcmin\\[3pt]
		 & 243$\pm$21 pc\\[3pt]
		\textbf{Ellipticity$^{c}$} & 0.67$\pm$0.03\\[3pt]
		\textbf{Position Angle$^{c}$} & -72.6$\pm$1.7$^{\circ}$\\[3pt]
		\textbf{Position Angle$^{e}$} & -78$\pm$4$^{\circ}$\\[3pt]
		\textbf{Stellar Mass}$^{d}$ & $3.7\times10^{4}$ M$_{\odot}$\\[3pt]
		\textbf{Luminosity$^{e}$} & $3.6\pm1.1\times$10$^{4}$ L$_{\odot}$\\[3pt]
		\hline
		\textbf{Heliocentric Velocity} & 46.4$\pm$1.3 kms$^{-1}$\\[3pt]
		\textbf{Velocity Dispersion} & 4.4$^{+1.4}_{-1.2}$ kms$^{-1}$\\[3pt]
		\textbf{Dynamical Mass (<1.8r$_{\rm{half}}$)} & 6.9$^{+4.4}_{-3.8}\times10^{6}$ M$_{\odot}$\\[3pt]
		\textbf{[Fe/H]} & $-2.48\pm0.19$ dex\\[3pt]
		\textbf{[Fe/H] Dispersion} & $0.63^{+0.18}_{-0.13}$ dex\\[3pt]
		\boldmath{$\mu_{\alpha}^*$} & -0.153$\pm$0.074 mas yr$^{-1}$\\[3pt]
		\boldmath{$\mu_{\delta}$} & -0.397$\pm$0.063 mas yr$^{-1}$\\[3pt]
		\textbf{Pericentre} & 50.9$^{+24.2}_{-23.6}$ kpc\\[3pt]
		\textbf{Apocentre} & 227.9$^{+85.1}_{-38.1}$ kpc\\[3pt]
		\textbf{Eccentricity} & 0.65$^{+0.10}_{-0.05}$\\[3pt]
		\hline
	\end{tabular}
	\label{tab:herc_properties}
\end{table}

\section{Observations and Data}\label{sec:obs}

\subsection{Photometry}
We utilise $B$-- and $r$-- band imaging of Hercules from the Large Binocular Telescope (LBT), first presented in \citet{sand09}. Observations were made in May and June 2008 using the Large Binocular Camera (LBC, \citealt{ragazzoni06}). LBT consists of two 8.4m telescopes, each equipped with a prime focus imager, with one optimised for blue wavelengths and one for red wavelengths. Each camera has a 23' $\times$ 23' field of view. Six 300s exposures were taken in each of the five fields, with seeing ranging from 0.8 to 1.4 arcsec. Full details of the observations and data reduction process are available in \citet{sand09}. Magnitudes were calibrated using stars in common with the Sloan Digital Sky Survey (SDSS, \citealt{adelman06}) DR6\footnote{\url{http://cas.sdss.org/DR6/en/}}, using the calibrations from \citet{jordi06} for the $B$-band. In cases where stars were near the saturation limit of LBT, the corresponding SDSS magnitudes are used instead.

\subsection{Spectroscopy}

The DEIMOS instrument \citep{faber03} on the Keck II telescope located on Mauna Kea in Hawaii was used to obtain spectroscopy of Hercules on 17th May 2015. DEIMOS is a multi--slit spectrograph with a field of view of approximately $16'\times14'$. We utilised the 1200 l/mm grating ($R\approx$6500), with a central wavelength of 7800 \AA\, and the OG550 filter to block shorter wavelengths. This is the same instrument configuration as used in SG07. Three DEIMOS slit masks were observed, designed to cover the field of view of SG07, as well as extending further along the major and minor axes to detect any tidal features. The approximate positions of these masks are shown in Fig. \ref{fig:obsmap}. Each mask contains 60--75 stars. Exposures of 1200--1800s were used, with total integration times ranging between 60 and 100 minutes per mask. Targets were selected using the LBT photometry. Typical seeing of between 0.7 and 1 arcsec was achieved in clear conditions.

Data reduction was performed following the process detailed in \citet{tollerud12,tollerud13}. The \textsc{spec2d} DEIMOS reduction pipeline was used to extract a one--dimensional spectrum for each target. The line--of--sight velocities are measured by cross correlating our spectra with high S/N spectra of known velocity stars, using either the H$_{\alpha}$ line or some aggregate of the weaker molecular lines. Mis--centring of stars within the slits is corrected for using strong telluric lines. Errors on the radial velocities are determined through a Monte Carlo process, outlined in \citet{tollerud12}. Random noise seeded by the variance per pixel is added to each spectrum, and the velocity determined by cross correlation. This process is repeated for 1000 iterations. The mean and standard deviation of these 1000 iterations are quoted as the radial velocity and error of the given target. Of the 205 observed targets, 173 were successfully reduced with velocity uncertainties <30 kms$^{-1}$. This sample has a mean S/N of 15.8 pix$^{-1}$ and a mean velocity uncertainty of $\delta_{\mathrm{v}}=2.5$ kms$^{-1}$.

\subsection{Defining Membership}\label{sec:members}

\begin{figure}
	\centering
	\includegraphics[width=\linewidth] {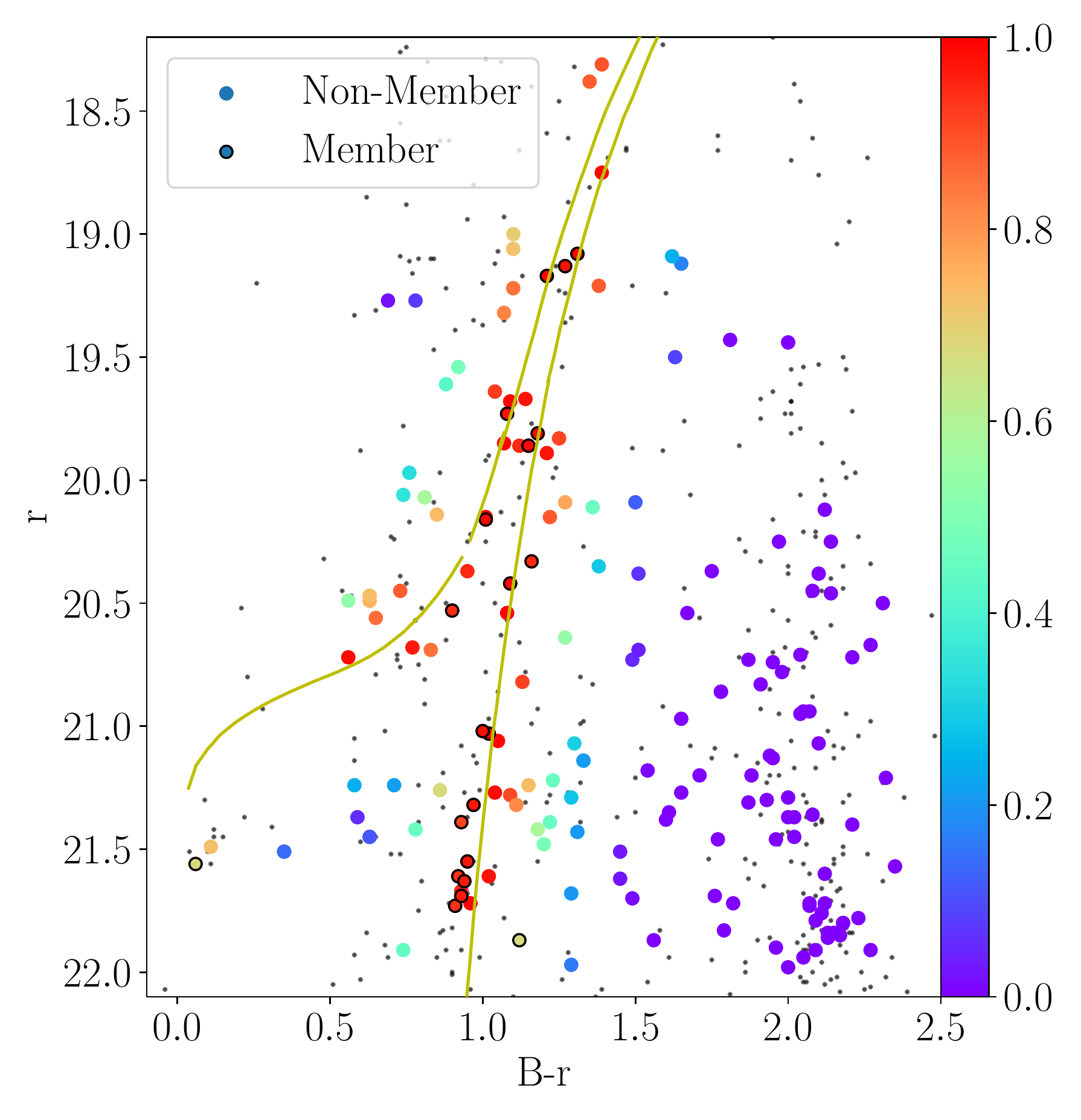}
	\caption{Colour--magnitude diagram for Hercules with a 12 Gyr, Z=0.0001 Padova isochrone overlaid. The targets observed with Keck/ DEIMOS are colour coded by their proximity to the isochrone, $P_{\mathrm{CMD}}$. Sources later confirmed as members are shown with a black outline.}
	\label{fig:isochrone}
\end{figure}

\begin{figure}
	\centering
	\includegraphics[width=\linewidth] {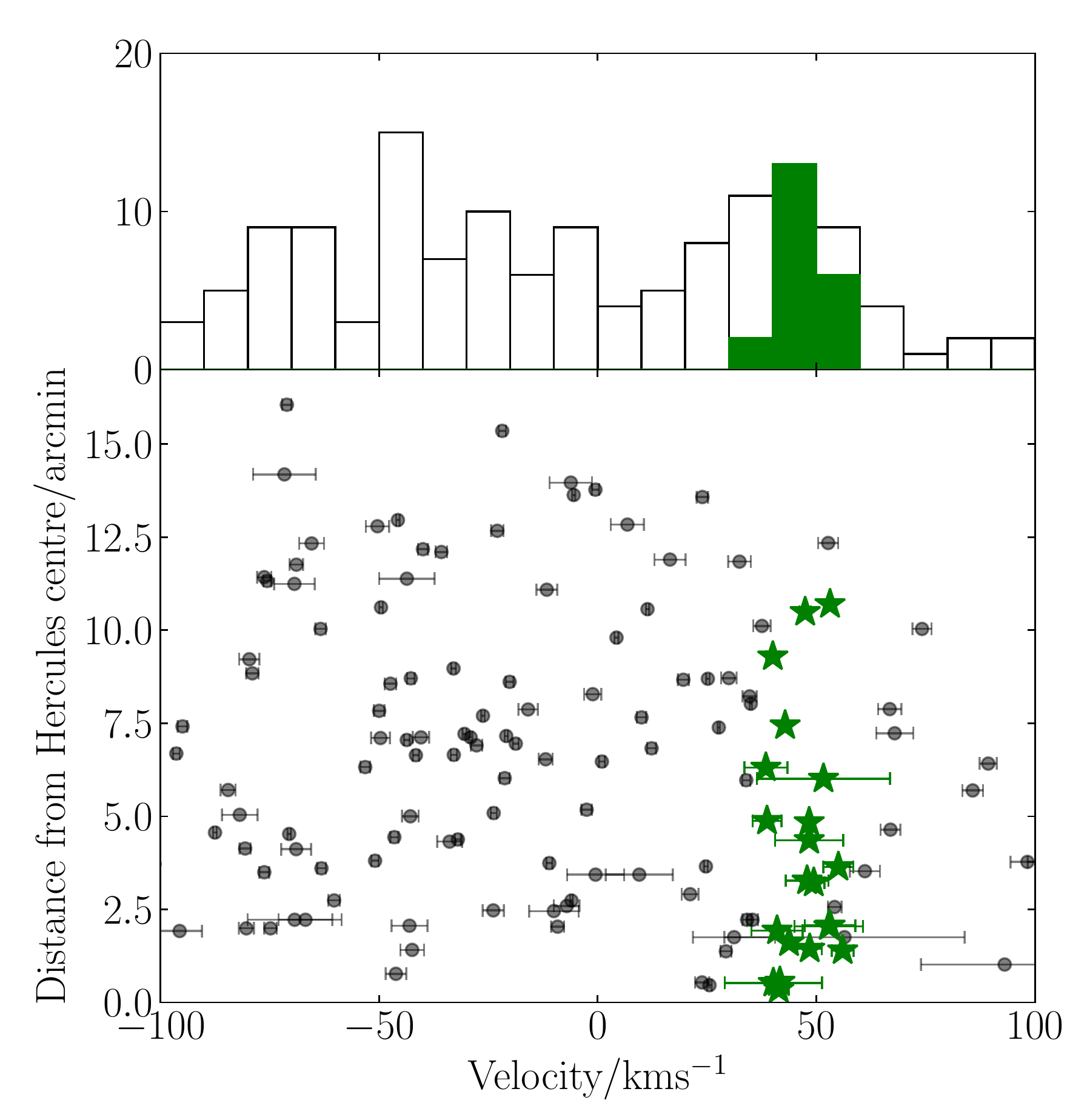}
	\caption{Top: Velocity histogram for the observed targets in Hercules, with those identified as members highlighted in green. Bottom: Measured stellar line--of--sight velocities as a function of distance from the centre of the Hercules dSph, again with those identified as members highlighted in green.}
	\label{fig:velrad}
\end{figure}

To classify the spectroscopic members of Hercules, we follow the probabilistic method outlined in \citet{collins13}. The probability of a star belonging to Hercules, given its position on the colour--magnitude diagram and its velocity, is 

\begin{eqnarray}
P_{i}\propto P_{\mathrm{CMD}} \times P_{\mathrm{vel}}.
\label{eq:probability}
\end{eqnarray}

To determine $P_{\mathrm{CMD}}$, we measure the position of each object on the colour--magnitude diagram (CMD) relative to a fiducial isochrone. We plot the CMD of Hercules using the photometric data from LBT in Fig. \ref{fig:isochrone}, taking only the sources within 7 arcmin of the centre to reduce crowding. We overlay two isochrones to best represent the horizontal branch (HB) and red giant branch (RGB) of Hercules. The isochrones are taken from the Padova database \citep{girardi02}, and describe a 12 Gyr, $Z=0.0001$ evolutionary track, shifted by a distance modulus $m-M=20.60$ to account for the distance to Hercules. \citet{tollerud12} define the distance from a given object to the isochrone as

\begin{eqnarray}
P\mathrm{_{CMD}}=\exp\bigg[-\frac{\Delta(B-r)^2}{2\sigma_c}-\frac{\Delta(r)^2}{2\sigma_m}\bigg],
\label{eq:pcmd}
\end{eqnarray}

\noindent where $\Delta(B-r)$ and $\Delta(r)$ are the minimum separation from the isochrone in each dimension, and $\sigma_c$ and $\sigma_m$ are free parameters accounting for distance and photometry factors. We set $\sigma_c=0.03$ and $\sigma_m=0.1$ to best constrain the Hercules HB and RGB. $P_{\mathrm{CMD}}$ serves as a proxy for the probability of membership. In Fig. \ref{fig:isochrone}, we plot the observed Hercules targets colour coded by $P_{\rm{CMD}}$.

Fig. \ref{fig:velrad} shows the histogram of radial velocities of our observed targets. The probability of a star belonging to a particular peak is given by \citep{collins13}

\begin{landscape}
\begin{table}
	\centering
	\caption{Details of the 21 identified members of Hercules as observed with Keck/DEIMOS. Columns are: (1) DEIMOS field (2) Object number; (3) Line--of--sight heliocentric velocity with velocity error; (4) Right Ascension in J2000; (5) Declination in J2000; (6) \textit{b}--band magnitude; (7) \textit{r}--band magnitude; (8) S/N ratio per pixel; (9) Distance from the centre of Hercules; (10) Metallicity; (11) Proper Motion in RA; (12) Proper Motion in Declination; (13) Position on CMD}
	\label{tab:members}
	\begin{tabular}{|c|c|c|c|c|c|c|c|c|c|c|c|c|}
		\hline
		\textbf{Field} & \textbf{Object} & \textbf{Velocity (kms$^{-1}$)} & \textbf{RA (deg)} & \textbf{Dec. (deg)} & \textbf{\textit{b}} & \textbf{\textit{r}} & \textbf{S/N (pix$^{-1}$)} & \textbf{Radius ('')} & \textbf{[Fe/H]} & \textbf{$\mu_{\alpha}^*$(mas)} & \textbf{$\mu_{\delta}$ (mas)} & \textbf{CMD} \\
		\hline
		H1 & 29  & 38.5$\pm$5.4  & 247.6827 & 12.7168 & 22.53 & 21.61 & 6.9  & 378.41 & -2.47$\pm$0.24 & -- & -- & RGB \\[3pt]
		& 60  & 47.9$\pm$5.4  & 247.7292 & 12.8383 & 22.32 & 21.39 & 8.5  & 196.32 & -- & -- & -- & RGB \\[3pt]
		& 89  & 41.4$\pm$3.1  & 247.7566 & 12.7861 & 22.50 & 21.55 & 7.7  & 21.06  & -- & -- & -- & RGB \\[3pt]
		& 100 & 40.2$\pm$11.3 & 247.7672 & 12.7923 & 22.64 & 21.73 & 6.3  & 31.36  & -2.78$\pm$0.45 & -- & -- & RGB \\[3pt]
		& 108 & 38.7$\pm$4.0  & 247.7714 & 12.8720 & 22.05 & 21.03 & 11.5 & 292.70 & -2.90$\pm$0.17 & -- & -- & RGB \\[3pt]
		& 115 & 48.5$\pm$3.5  & 247.7758 & 12.8086 & 22.99 & 21.87 & 5.6  & 86.57  & -- & -- & -- & RGB \\[3pt]
		& 117 & 53.1$\pm$6.2  & 247.7777 & 12.8204 & 22.62 & 21.69 & 6.4  & 123.81 & -1.26$\pm$0.28 & -- & -- & RGB \\[3pt]
		 \hline
		H2 & 5   & 42.9$\pm$2.2  & 247.8740 & 12.7403 & 20.38 & 19.17 & 42.4 & 446.55 & -2.57$\pm$0.03 & -0.153$\pm$0.369 & -0.134$\pm$0.306 & RGB \\[3pt]
		& 8 & 56.0$\pm$3.3 & 247.7687 & 12.7707 & 20.99 & 19.81 & 9.8 & 83.81 & -- & 0.551$\pm$0.640 & 0.663$\pm$0.528 & RGB \\[3pt]
		& 14 & 52.9$\pm$8.2 & 247.7901 & 12.7777 & 22.57 & 21.63 & 4.8 & 122.62 & -- & -- & -- & RGB \\[3pt]
		& 23  & 55.0$\pm$4.1  & 247.8168 & 12.7712 & 22.02 & 21.02 & 11.3 & 218.25 & -2.69$\pm$0.17 & -- & -- & RGB \\[3pt]
		& 41 & 51.6$\pm$15.4 & 247.8610 & 12.7930 & 21.62 & 21.56 & 4.5 & 360.57 & -- & -- & -- & HB\\[3pt]
		& 56  & 40.1$\pm$2.6  & 247.9092 & 12.7435 & 21.17 & 20.16 & 19.8 & 557.67 & -2.88$\pm$0.08 & -1.529$\pm$0.794 & -0.526$\pm$0.680 & RGB \\[3pt]
		\hline
		H3 & 2   & 43.8$\pm$2.3  & 247.7838 & 12.8017 & 20.39 & 19.08 & 27.2 & 96.63  & -- & -0.457$\pm$0.362 & -0.620$\pm$0.276 & RGB \\[3pt]
		& 7   & 48.4$\pm$2.3  & 247.6854 & 12.8299 & 20.40 & 19.13 & 34.0 & 290.54 & -2.73$\pm$0.03 & 0.541$\pm$0.378 & -0.521$\pm$0.288 & RGB \\[3pt]
		& 30  & 47.5$\pm$2.6  & 247.5934 & 12.8603 & 20.81 & 19.73 & 22.9 & 629.26 & -3.02$\pm$0.08 & -0.116$\pm$0.591 & -0.486$\pm$0.461 & RGB \\[3pt]
		& 35  & 53.1$\pm$2.5  & 247.6033 & 12.8864 & 21.49 & 20.33 & 13.8 & 641.92 & -1.31$\pm$0.13 & -- & -- & RGB \\[3pt]
		& 67  & 48.4$\pm$8.1  & 247.6888 & 12.8180 & 22.29 & 21.32 & 5.8  & 261.56 & -1.85$\pm$0.31 & -- & -- & RGB \\[3pt]
		& 73  & 49.4$\pm$3.3  & 247.7041 & 12.8004 & 21.43 & 20.53 & 12.0 & 192.89 & -3.28$\pm$0.23 & 0.003$\pm$1.108 & 0.057$\pm$0.911 & RGB \\[3pt]
		& 95  & 41.7$\pm$2.5  & 247.7600 & 12.8008 & 21.01 & 19.86 & 19.6 & 33.22  & -2.40$\pm$0.06 & -0.495$\pm$0.670 & -0.764$\pm$0.518 & RGB \\[3pt]
		& 109 & 41.0$\pm$6.2 & 247.7911 & 12.7951 & 21.51 & 20.42 & 7.6 & 115.74 & -- & -0.322$\pm$1.050 & -0.043$\pm$0.801 & RGB \\[3pt]
		\hline              
	\end{tabular}
\end{table}
\end{landscape}

\begin{multline}
P_{\mathrm{peak},i}=\frac{1}{\sqrt{2\pi}\sqrt{\sigma_{v,\mathrm{peak}}^2+v_{\mathrm{err}, i}^2}} \\ 
\times\exp\bigg(-\frac{1}{2}\bigg[\frac{v\mathrm{_{peak}}-v_i}{\sqrt{\sigma_{v,\mathrm{peak}}^2+v_{\mathrm{err}, i}^2}}\bigg]^2\bigg),
\label{eq:probpeak}
\end{multline}

\noindent where $v_i$ is the velocity of the given star, $v_{\mathrm{err}}$ is the error on its velocity, $v_{\mathrm{peak}}$ is the systemic velocity of the peak, and $\sigma_{v,\mathrm{peak}}$ is the velocity dispersion. An additional systematic uncertainty of 2.2kms$^{-1}$ is known to be present in DEIMOS spectra (SG07; \citealt{kalirai10,tollerud12}); this is added in quadrature to our measured velocity uncertainty. We then use Bayesian inferences to define the probability of a star being a velocity member as

\begin{eqnarray}
P\mathrm{_{vel}}=\frac{P_{\mathrm{Herc},i}}{P_{\mathrm{MW},i} + P_{\mathrm{Herc},i}}.
\label{eq:pvel}
\end{eqnarray} 

Based on velocity alone, it is difficult to differentiate between the Milky Way foreground and Hercules members as they occupy the same velocity space. This results in heavy foreground contamination of the Hercules peak (see Fig. \ref{fig:velrad}), vastly inflating the velocity dispersion and skewing the measured velocity. We therefore define the Hercules peak using the values measured by A09, with $v_{\mathrm{peak}}=45.20$ kms$^{-1}$ and $\sigma_{v,\mathrm{peak}}=3.72$ kms$^{-1}$. These values are chosen because A09 were able to utilise intermediate band Str\"{o}mgren photometry to robustly separate the foreground stars from the dSph members, meaning the systemic velocity and velocity dispersion were measured from a well defined population. For the Milky Way peak, we define $v_{\mathrm{peak}}=-10.0$ kms$^{-1}$ and $\sigma_{v,\mathrm{peak}}=60$ kms$^{-1}$, values estimated from a Besan\c{c}on model \citep{robin03} of the Milky Way in the Hercules region of the sky.

$P_{\mathrm{CMD}}$ and $P_{\mathrm{vel}}$ are combined in equation \ref{eq:probability} to obtain the overall probability of membership for each observed target. Note that we do not include a factor to account for the distance from the object to the centre of Hercules. The DEIMOS masks were specifically designed to lie across the centre of the galaxy, and to exclude the more distant objects may inhibit our ability to observe tidal features. We define the members of Hercules as those stars with $P_i>0.5$. This results in a population of 21 member stars from this dataset, comprising 20 RGB and 1 HB star. The velocities of the members are plotted as a histogram and as a function of distance from the centre of Hercules in Fig. \ref{fig:velrad}, and are highlighted in Fig. \ref{fig:isochrone}. Our sample of members has a mean S/N of 13.7pix$^{-1}$ and a mean velocity uncertainty of $\delta_{\mathrm{v}}=5.0$ kms$^{-1}$.

\subsubsection{Tidally Stripped Members?}

Our field of view is focussed on the central regions of Hercules; hence the majority of our member stars fall inside the half--light radius. There are five members of our sample located outside this radius (see Fig. \ref{fig:obsmap}). Comparing to the overdensities identified in \citet{roderick15}, these stars may fall in the contour surrounding the central $r_{\rm{half}}$ (`Segment 13'), and so may represent material which has been tidally stripped. This tentative detection appears to suggest that the debris surrounding Hercules is indeed associated with the galaxy. However, we note that these stars still fall within 2 $r_{\rm{half}}$, and may be bound stars which have been preferentially observed due to the positioning of the DEIMOS slit masks. Either way, these and further flung regions would be strong candidates for further spectroscopic study.

\section{Chemo-Dynamics of Hercules}\label{sec:chemodyn}

\subsection{Kinematics}\label{sec:kinematic}

\subsubsection{Dispersion Supported System} \label{sysvel}

We first consider the scenario where Hercules is a bound, non--disrupting system with little or no rotation. To determine the systemic velocity and velocity dispersion, we follow the maximum likelihood method first defined in \citet{martin07}. An MCMC routine is used to find the velocity and velocity dispersion which maximise the log--likelihood function

\begin{eqnarray}
\log \mathcal L=\sum_{i=1}^{N} \log(\eta\mathrm{_{Herc}} P_{\mathrm{Herc},i}),
\label{eq:likelihood}
\end{eqnarray}

\begin{figure}
	\centering
	\includegraphics[width=\linewidth] {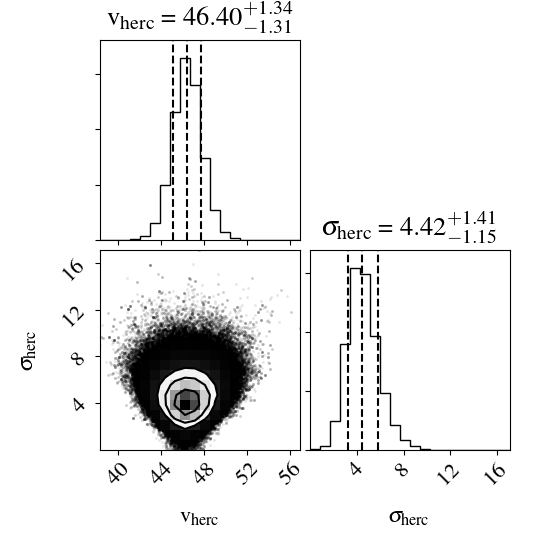}
	\caption{Two--dimensional and marginalized PDFs for the systemic velocity and velocity dispersion (both in kms$^{-1}$) of the identified Hercules members, assuming a purely dispersion supported system. The dashed lines represent the mean value and 1$\sigma$ uncertainties.}
	\label{fig:emcee_gaussian}
\end{figure}

\noindent where $P_{\mathrm{Herc},i}$ is the probability of belonging to the Hercules velocity peak and $\eta$ defines the fraction of the total population belonging to each peak. $P_{\mathrm{Herc},i}$ is calculated from equation \ref{eq:probpeak}, multiplied by an additional factor $P_i$ to account for the probability of membership of the given object. We apply flat priors such that 20 kms$^{-1}<v_{\mathrm{Herc}}<75$kms$^{-1}$ and 0 kms$^{-1}<\sigma_{v,\mathrm{Herc}}<20$ kms$^{-1}$. The \textsc{emcee} sampler \citep{foremanmackey13} is used to explore this parameter space. Fig. \ref{fig:emcee_gaussian} shows the resulting 1D and 2D probability distribution functions for the parameters $v_{\mathrm{Herc}}$ and $\sigma_{v,\mathrm{Herc}}$. The resulting values are $v_{\mathrm{Herc}}=46.4\pm1.3$ kms$^{-1}$ and $\sigma_{v,\mathrm{Herc}}=4.4^{+1.4}_{-1.2}$ kms$^{-1}$, where the quoted uncertainties are the 1--$\sigma$ (68th percentile) confidence bounds. These results are fully consistent with the velocity and velocity dispersion results of SG07 and A09.

Under the assumption that Hercules is a dispersion supported system, we can calculate the dynamical mass using the relation defined by \citet{errani18}, where

\begin{eqnarray}
M(<1.8r_{\rm{half}})=\frac{1.8 \mu r\mathrm{_{half}} \sigma_{v,\mathrm{half}}^{2}}{G}.
\label{eq:mass}
\end{eqnarray}

\noindent $r_{\mathrm{half}}$ is the half light radius and $\mu=3.5$. This mass estimator is chosen because it does not assume a flat dispersion profile. Using our measured value for $\sigma_{v,\mathrm{Herc}}$, and taking $r_{\mathrm{half}}=243$ pc \citep{sand09}, this returns a mass of $M(<1.8r_{\rm{half}})=6.9^{+4.4}_{-3.8}\times10^{6}$ M$_{\odot}$, corresponding to a mass--to--light ratio of $M/L(<1.8r_{\rm{half}})\approx191^{+123}_{-105}$  M$_{\odot}$/L$_{\odot}$. This would imply a strongly dark matter dominated system.

\subsubsection{Disrupting System}

\begin{table*}
	\centering
	\caption{Kinematics of Hercules as measured in this work. The velocity gradient results are listed with the position angle fixed along the major axis ($\theta$=-78$^{\circ}$); along the minor axis ($\theta$=+12$^{\circ}$); and allowed to converge freely.}
	\begin{tabular}{|l|c|c|c|c|}
		\hline
		 & \textbf{No Gradient} & \textbf{Major Axis} & \textbf{Minor Axis} & \textbf{Free Position Angle}\\ \hline
		\textbf{Velocity} & $46.4\pm1.3$ kms$^{-1}$ & $46.1^{+1.3}_{-1.2}$ kms$^{-1}$ & $46.1\pm1.4$ kms$^{-1}$ & $46.2\pm1.4$ \\[3pt]
		\textbf{Velocity Dispersion} & $4.4^{+1.4}_{-1.2}$ kms$^{-1}$ & $4.5^{+1.4}_{-1.1}$ kms$^{-1}$ & $5.1^{+1.3}_{-1.0}$ kms$^{-1}$ & $4.7^{+1.4}_{-1.1}$ kms$^{-1}$ \\[3pt]
		\textbf{Velocity Gradient} & -- & $9.2^{+5.9}_{-6.3}$ kms$^{-1}$kpc$^{-1}$  & $9.0^{+19.5}_{-19.9}$ kms$^{-1}$kpc$^{-1}$ & $9.5^{+9.8}_{-13.0}$ kms$^{-1}$kpc$^{-1}$ \\[3pt]
		 & -- & $20.9^{+13.6}_{-14.4}$ kms$^{-1}$deg$^{-1}$ & $18.2^{+40.6}_{-43.5}$ kms$^{-1}$deg$^{-1}$ & $21.8^{+22.2}_{-29.9}$ kms$^{-1}$deg$^{-1}$ \\[3pt]
		 \textbf{Position Angle of Gradient} & -- & -72.6$^{\circ}$ & +17.4$^{\circ}$ & $-95.7^{\circ}$$^{+68.9}_{-56.9}$ \\
		 \hline
	\end{tabular}
	\label{tab:herc_kinematics}
\end{table*}

Previous photometric and spectroscopic studies indicate some form of disruption within Hercules (\citealt{coleman07,sand09,martin10,roderick15}, among others). We therefore consider a second scenario, and search for kinematic evidence of this tidal disturbance in the form of a velocity gradient. We follow the maximum likelihood method laid out in \citet{martin10} to find the velocity gradient $\frac{dv_{r}}{d\chi}$ acting along an axis with position angle $\theta$. The likelihood function defined in equation \ref{eq:likelihood} is modified to become

\begin{eqnarray}
\mathcal{L}\bigg(v,\sigma_v,\frac{dv_{r}}{d\chi},\theta\bigg)=\prod_i l_i \bigg(v,\sigma_v,\frac{dv_{r}}{d\chi},\theta\bigg)
\label{eq:gradlikelihood}
\end{eqnarray}

\noindent where

\begin{eqnarray}
l_i=\frac{1}{\sqrt{2\pi}\sqrt{\sigma_{v, \mathrm{Herc}}^2+v_{\mathrm{err}, i}^2}} \times  \exp\bigg(-\frac{1}{2}\bigg[\frac{\Delta v_{r,i}^{2}}{\sqrt{\sigma_{v, \mathrm{Herc}}^2+v_{\mathrm{err}, i}^2}}\bigg]^2\bigg)
\label{eq:gradient}
\end{eqnarray}

\noindent describes the probability of finding the object $i$ given the parameters. $\Delta v_{r,i}$ is the difference between the measured velocity of a star and a velocity gradient $\frac{dv_{r}}{d\chi}$ acting along the angular distance of a star along an axis $y_i$ with position angle $\theta$, and is given by

\begin{eqnarray}
\Delta v_{r,i}=v_{r,i}-\frac{dv_{r}}{d\chi}y_{i}+\bar{v_{r}}
\label{eq:deltav}
\end{eqnarray}

\noindent $y_i$ is defined by the RA and Dec. of the star, ($\alpha_i$,$\delta_i$), relative to the coordinates of the centre of Hercules, ($\alpha_0$,$\delta_0$), as listed in Table \ref{tab:herc_properties}.

\begin{eqnarray}
y_i&=&X_{i}\sin{\theta}+Y_{i}\cos{\theta}\\
X_i&=&(\alpha_{i}-\alpha_{0})\cos(\delta_{0})\\
Y_i&=&\delta_{i}-\delta_{0}
\label{eq:gradeqns}
\end{eqnarray}

We again use \textsc{emcee} to implement an MCMC routine and find the best fit set of parameters. We use the same priors as in the original model, and introduce priors on the gradient and position angle such that -100 kms$^{-1}$ kpc$^{-1}<\frac{dv_{r}}{d\chi}<$ 100kms$^{-1}$ kpc$^{-1}$ and $-\pi<\theta<0$, where $\theta$ is measured from North to East. 

\begin{figure}
	\centering
	\includegraphics[width=\linewidth] {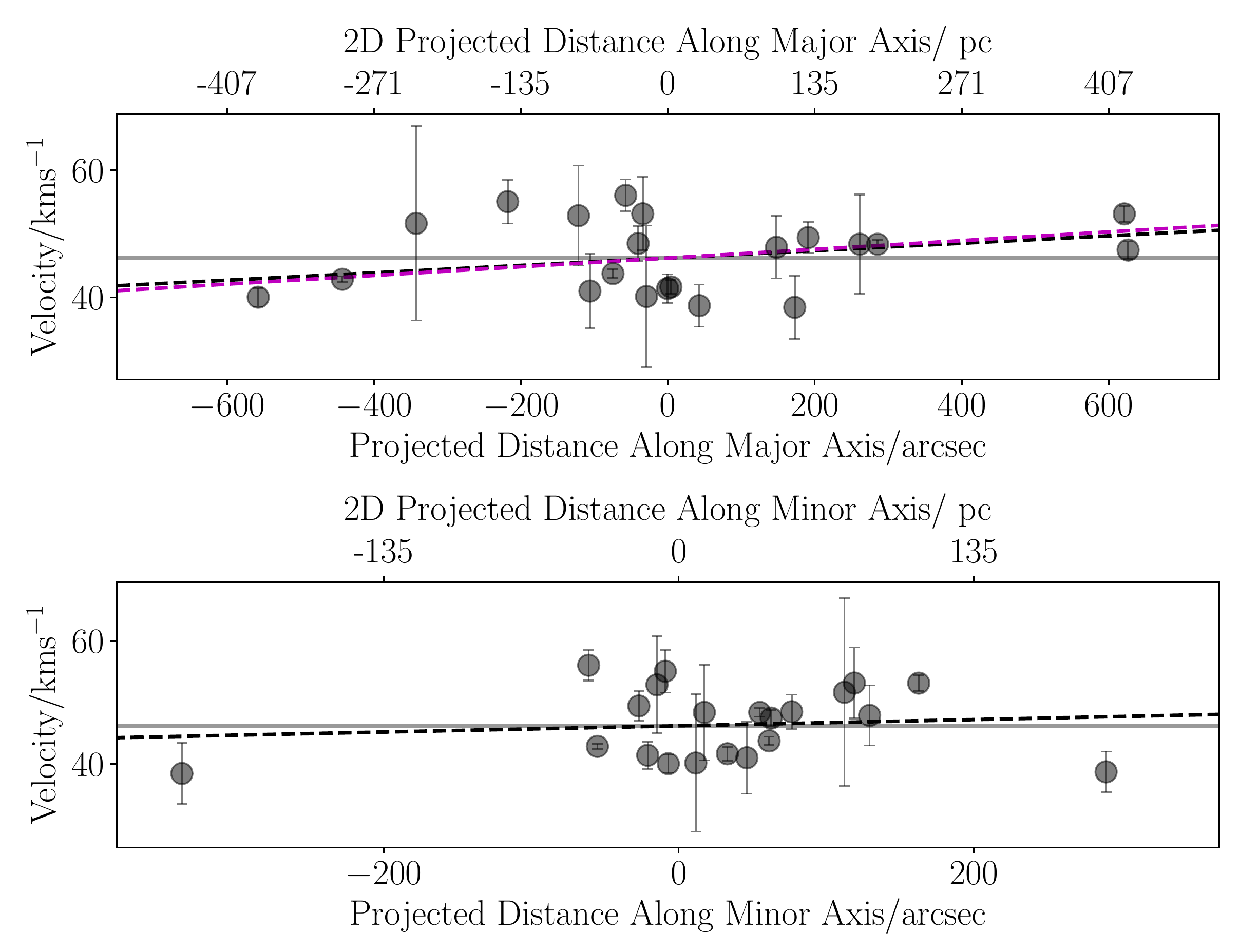}
	\caption{Stellar velocities plotted as a function of position along the major and minor axes of Hercules. The black dashed line highlights the velocity gradient measured in this work. The dashed magenta line shows the velocity gradient calculated in \citet{martin10} from the velocities identified in \citet{aden09a}. The grey line marks the measured systemic velocity of Hercules.}
	\label{fig:velmajorax}
\end{figure}

If all parameters are allowed to evolve freely we measure a systemic velocity of  $v_{\mathrm{Herc}}=46.2\pm1.3$ kms$^{-1}$ and a velocity dispersion of  $\sigma_{v,\mathrm{Herc}}=4.7_{-1.1}^{+1.4}$ kms$^{-1}$. These are fully consistent with the result with no gradient. The velocity gradient is measured to be $\frac{dv_{r}}{d\chi}=9.5^{+9.8}_{-13.0}$ kms$^{-1}$kpc$^{-1}=21.8^{+22.2}_{-29.9}$ kms$^{-1}$deg$^{-1}$. These results do not provide strong support for the presence of a velocity gradient across the galaxy, and are consistent with zero within 1$\sigma$. As such, Hercules may not be as dynamically unstable as previously suggested.

\citet{martin10} note that if Hercules is a stellar stream in formation, the velocity gradient must be aligned with the major axis, such that the stars are flowing along the orbit of the progenitor. If we follow this reasoning, and fix $\theta=-72.6^{\circ}$ (the value from \citealt{sand09}) accordingly, we obtain a velocity gradient of $\frac{dv_{r}}{d\chi}=9.2^{+5.9}_{-6.3}$ kms$^{-1}$kpc$^{-1}=20.9^{+13.6}_{-14.4}$ kms$^{-1}$deg$^{-1}$. This result is consistent with \citet{martin10}, as shown in Fig. \ref{fig:velmajorax}, where we plot our stellar velocities as a function of position along the major axis. It is similar in magnitude to our `free' gradient result, but with smaller errors. We note that if we use the position angle of $-78^{\circ}$ from \citet{martin08}, as is used in \citet{martin10}, we obtain consistent results ($\frac{dv_{r}}{d\chi}=9.4^{+6.0}_{-6.3}$ kms$^{-1}$kpc$^{-1}=21.4^{+13.8}_{-14.7}$ kms$^{-1}$deg$^{-1}$).

On the other hand, in the `exploding satellite' scenario proposed by K17, Hercules is dissolving into a stream perpendicular to its orbit. In this case one would expect a velocity gradient aligned with the minor axis. Fixing $\theta=+17.4^{\circ}$ returns $\frac{dv_{r}}{d\chi}=9.0^{+19.5}_{-19.9}$ kms$^{-1}$kpc$^{-1}=18.2^{+40.6}_{-43.5}$ kms$^{-1}$deg$^{-1}$. Once again, the errors are large and would be consistent with zero gradient; therefore, we find no strong evidence in favour of a gradient across the minor axis.

The measured kinematics of Hercules are presented in Table \ref{tab:herc_kinematics}, under the assumptions of both a dispersion supported and a disrupting system.

\subsubsection{Comparison to Previous Studies}

\begin{figure}
	\centering
	\includegraphics[width=\linewidth] {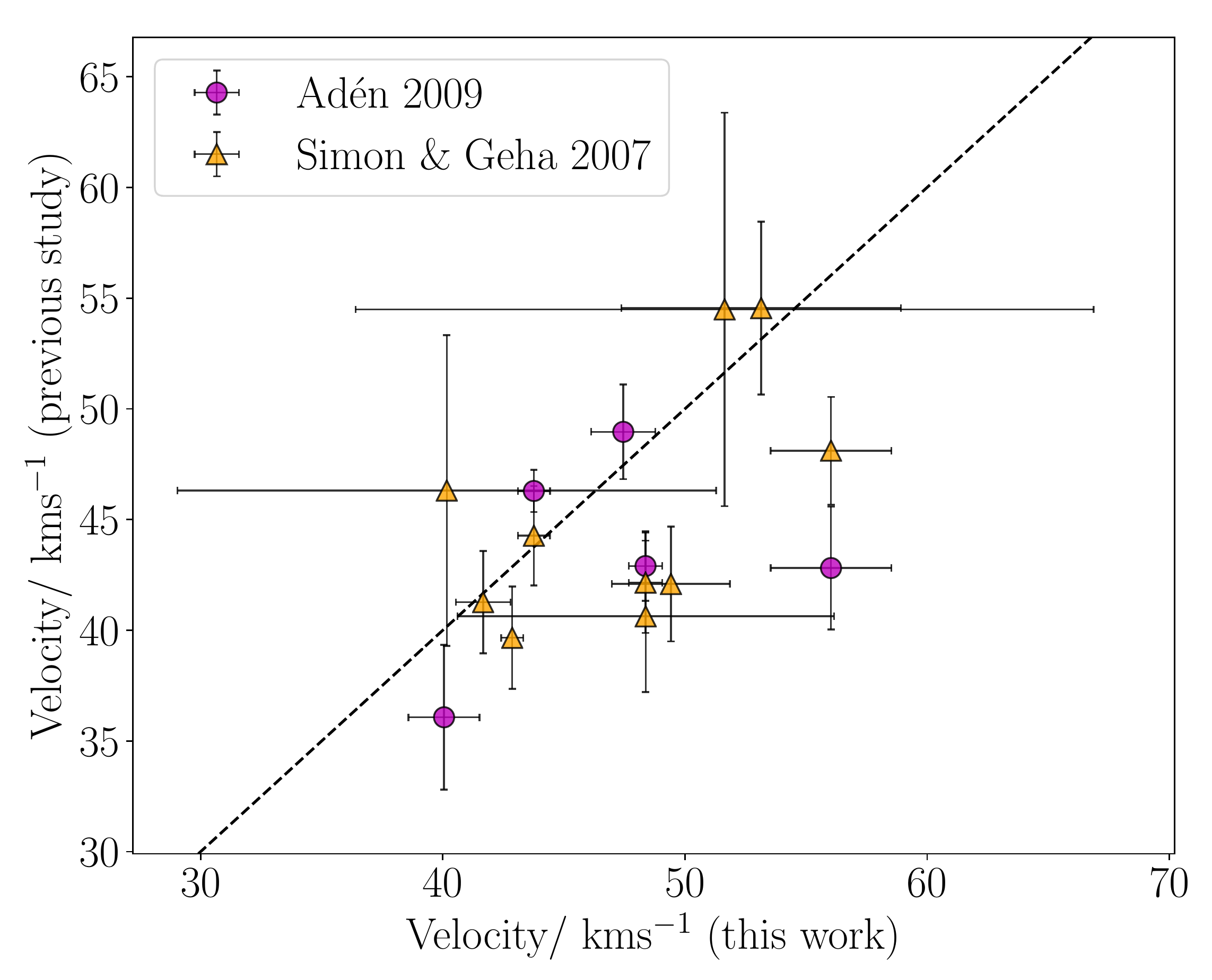}
	\caption{Comparison between the radial velocities of the members identified in this work and those identified in \citet{simon07} (orange triangles) and \citet{aden09a} (magenta circles). Some stars are present in all three datasets.}
	\label{fig:velcomparison}
\end{figure}

The kinematics of Hercules have previously been measured by both SG07 and A09. SG07 also used Keck/ DEIMOS observations of Hercules, obtaining 30 member stars in the plane of the galaxy. From these, they measure a radial velocity of $v_{\mathrm{Herc}}=45.0\pm1.1$ kms$^{-1}$ and a velocity dispersion of $\sigma_{v,\mathrm{Herc}}=5.1\pm0.9$ kms$^{-1}$, values which are fully consistent with our results. Our membership sample contains 10 stars in common with the members identified in SG07. The radial velocities of these stars are plotted in Fig. \ref{fig:velcomparison}. Six of the radial velocity measurements agree within 1--$\sigma$, and all agree within 3--$\sigma$. The differences between the velocity measurements of all 29 stars in common (including non--members) follow a normal distribution. Using only the members in common, from the velocities measured in this work, we obtain a systemic velocity measurement of $v_{\mathrm{Herc}}=46.9^{+2.1}_{-1.9}$ kms$^{-1}$ and a velocity dispersion of $\sigma_{v,\mathrm{Herc}}=4.5^{+2.4}_{-1.8}$ kms$^{-1}$. From the SG07 velocities, we measure a systemic velocity of $v_{\mathrm{Herc}}=44.1^{+1.6}_{-1.5}$ kms$^{-1}$ and a velocity dispersion of $\sigma_{v,\mathrm{Herc}}=2.6^{+2.4}_{-1.7}$ kms$^{-1}$.  These results are consistent with each other within 1--$\sigma$, and although the velocity dispersion appears small when using the SG07 velocities, this is probably due to the small sample size.

A09 use Str\"{o}mgren photometry to constrain a population of 47 Hercules members, and obtain kinematics for 20 of them using the FLAMES instrument on the VLT. From the 18 RGB members, they determine a radial velocity of $v_{\mathrm{Herc}}=45.20\pm1.09$ kms$^{-1}$ and a velocity dispersion of $\sigma_{v,\mathrm{Herc}}=3.72\pm0.91$ kms$^{-1}$. This velocity dispersion is slightly smaller than both the SG07 result and our measurement in this work. Comparing our dataset with that of A09, we find 5 overlapping stars. Two of these velocities, which are also shown in Fig. \ref{fig:velcomparison}, are consistent within 1--$\sigma$, and all are consistent within 3--$\sigma$. Using only these stars in common, from the Keck velocities (this work), we obtain a systemic velocity measurement of $v_{\mathrm{Herc}}=46.9^{+3.2}_{-3.1}$ kms$^{-1}$ and a velocity dispersion of $\sigma_{v,\mathrm{Herc}}=6.3^{+4.4}_{-2.6}$ kms$^{-1}$. From the A09 velocities, we obtain a a systemic velocity measurement of $v_{\mathrm{Herc}}=44.0^{+2.3}_{-2.5}$ kms$^{-1}$ and a velocity dispersion of $\sigma_{v,\mathrm{Herc}}=4.0^{+4.0}_{-2.5}$ kms$^{-1}$. These results are consistent within 1--$\sigma$.

In the analysis of their modelled orbit, K17 propose an observational test of the model whereby a velocity substructure should be detected between 41--43kms$^{-1}$. This is formed by the so-called `exploded component', that is, the stars aligned with the major axis and perpendicular to the orbital path, which should retain the velocity of the progenitor, and which also should not display a velocity gradient. We find no evidence for a velocity substructure in our membership sample, with only 4 of the 21 stars falling within the defined bounds. However, the average velocity uncertainty per member star in our dataset is $\delta_v=4.2$ kms$^{-1}$, which is comparable to the overall velocity dispersion and so may blur out any residual velocity substructures (see the discussion in Section 7 of K17 for more detail).

\subsection{Metallicity}\label{sec:feh}

The S/N of our spectra is sufficient to obtain individual metallicities of our member stars. To do this we utilise the empirical relation between the equivalent widths of the three Ca II absorption lines ($\lambda$=8498\AA, 8542\AA, 8662\AA), and the metallicity [Fe/H] of an RGB star. There are several calibrations available for this (e.g. \citealt{starkenburg10,carrera13}), which take the general form 

\begin{eqnarray}
\mathrm{[Fe/H]} = a + b M_V + c \mathrm{EW} + d \mathrm{EW}^{-1.5} + e \mathrm{EW} \times M_V,
\label{eq:metallicity}
\end{eqnarray}

\noindent for an RGB star of absolute $V$--band magnitude $M_V$. We choose to use the calibration defined in \citet{carrera13}, which is derived from observations of real stars and spans the metallicity range -4.0$<$[Fe/H]$<$0.5. In this case, the coefficients of equation \ref{eq:metallicity} are

\begin{eqnarray*}
a=-3.45 \; b=0.16 \; c=0.41 \; d=-0.53 \; e=0.019.
\label{eq:c13coefficients}    
\end{eqnarray*}

\noindent The absolute magnitude of a given Hercules member star is calculated from the apparent $V$--band magnitude, $m_V$, and the distance to Hercules using

\begin{eqnarray}
M_{V}=m_{V}-5\log_{10}(r_{\rm{Herc}})+5
\label{eq:magnitude}    
\end{eqnarray}

\noindent where it is assumed that all targets lie at the given distance of Hercules. The small error introduced by this assumption is outweighed by the error generated from the noise in the spectra.

Our spectra are processed as follows. We first normalise our spectra by smoothing with a median filter to fit to the continuum, and dividing the spectrum through by this continuum. We then fit a model consisting of a smooth continuum and three Gaussian peaks representing the Ca II lines to the normalised spectra using a least squares minimisation. The equivalent width of each peak is then extracted from the fitted peak area and continuum. This is repeated for all three lines, and the results are fed into equation \ref{eq:metallicity}.

We exclude our HB star from this analysis as the Ca II--metallicity relation is only calibrated for the RGB. From the RGB population, we obtain metallicities ranging from [Fe/H]= -1.26 dex to [Fe/H]= -3.28 dex for 13 stars. In some cases the spectrum was incomplete in the Ca II region, or was too noisy to produce a reliable metallicity measurement. The normalised Hercules member spectra used, with the Gaussian fit displayed in black, are presented in appendix \ref{appendixa}. Using \textsc{emcee} to fit a Gaussian to the histogram of metallicities, we obtain a mean metallicity of [Fe/H]= $-2.48\pm0.19$ dex and a dispersion of $\sigma_{\rm{[Fe/H]}}= 0.63^{+0.18}_{-0.13}$ dex. The individual metallicities of member stars are listed in table \ref{tab:members}. Errors on these values are obtained by propagating the values from the covariance matrix of the fitting parameters.

\begin{figure}
	\centering
	\includegraphics[width=\linewidth] {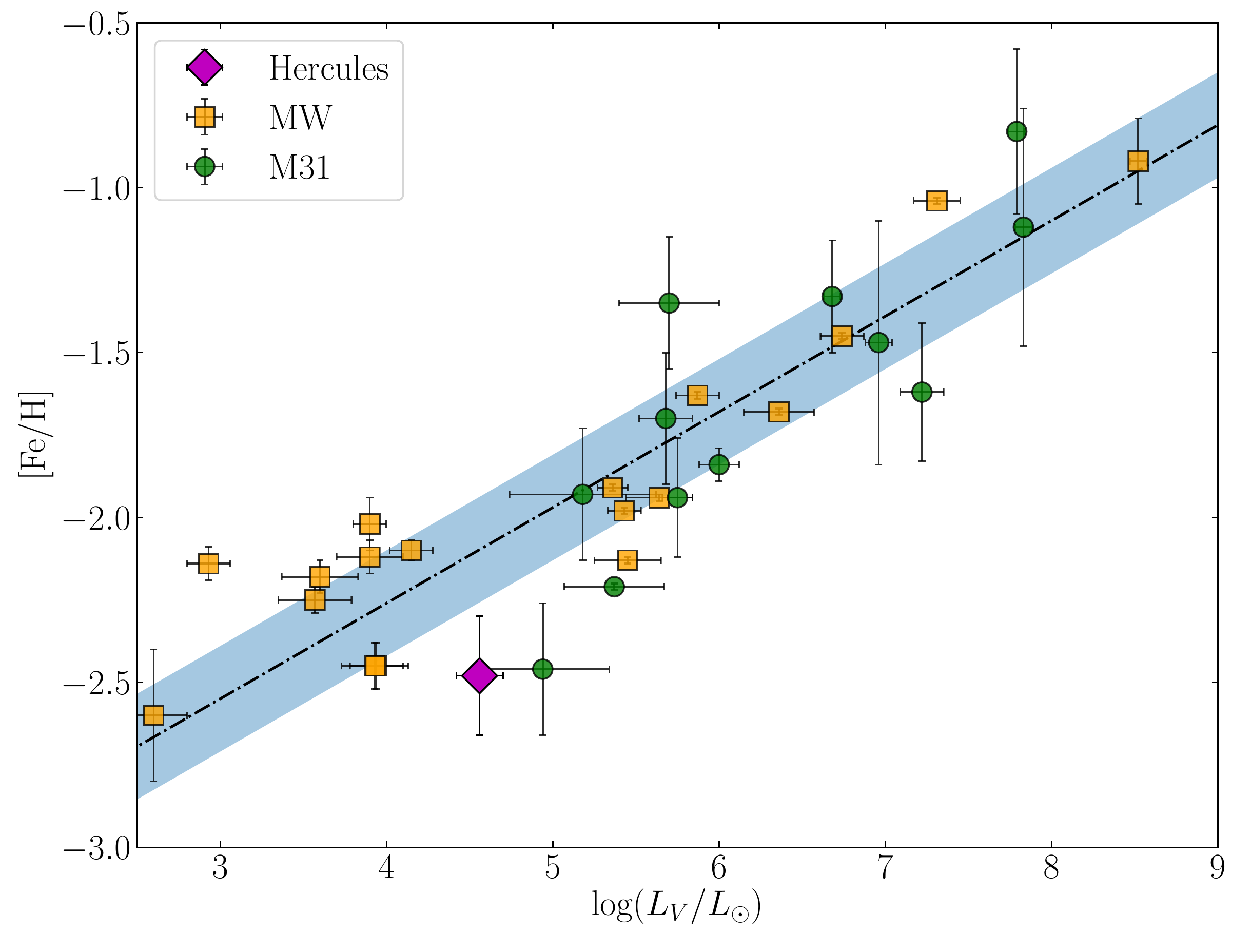}
	\caption{Metallicity as a function of log ($L_V/L_{\odot}$) for the Milky Way (orange squares) and M31 (green points) dwarfs. Hercules is shown as a magenta diamond. The [Fe/H]--$L_V$ relation defined in \citet{kirby13} is plotted as a dot--dash line, with the rms scatter about this relation shown as the shaded region. Our measured metallicity places Hercules $\sim$1--$\sigma$ below this relation. Galaxy data mostly from \citet{kirby13}, with additional data points from \citet{kirby15,martin16,longeard18}.}
	\label{fig:fehlumplot}
\end{figure}

This result is in agreement with previous findings that Hercules is a very metal--poor galaxy. Our result finds it to be slightly more metal poor than A09, who obtained a photometric metallicity of [Fe/H]=$-2.35\pm0.35$ dex, and SG07, who measured [Fe/H]=$-2.27\pm0.07$ dex. Perhaps the most robust measurement thus far is by \citet{kirby08a}, who measured a metallicity of [Fe/H]=$-2.58\pm0.04$ dex by comparing to synthetic spectra of RGB stars (see \citealt{kirby08b}). Our mean metallicity is within 1--$\sigma$ of this measurement. \citet{kirby08a} measure a metallicity dispersion of $\sigma_{\rm{[Fe/H]}}= 0.51$ dex; again, our result is within 1--$\sigma$ of this measurement. The large spread of metallicities is not uncommon in Milky Way dwarfs \citep{kirby13}, and indicates self--enrichment in the star formation history of the galaxy. Under the assumption that a deep gravitational well is required to retain enrichment products, this confirms that the progenitor of the Hercules stream was (is-- if not yet unbound) a dwarf galaxy.

%\begin{figure}
%	\centering
%	\includegraphics[width=\linewidth] {fehgradient_final.pdf}
%	\caption{Metallicity as a function of distance from the centre of Hercules for our member stars. We detect a slight negative gradient when considering all stars ( line). If we exclude the high metallicity, most distant member (diamond marker), the gradient increases to $\frac{d\rm{[Fe/H]}}{d\rm{r}}=-0.045\pm0.043$ dex arcmin$^{-1}$ ( line).}
%	\label{fig:fehgradient}
%\end{figure}

Using the normalised spectra, we can perform one final test of our defined membership sample by checking the Na I doublet absorption feature. The Ca II--[Fe/H] relation, which does not apply to dwarf stars, nonetheless may produce low metallicity results for Milky Way interlopers. Given the large spread in measured metallicities, it is therefore possible that some contaminants are still present. We check this by examining the Na I doublet, centred on $\sim$8200\AA, which is typically much stronger in dwarf stars than in giants. Indeed, we find no absorption features at this location in any of our member spectra, increasing confidence in our sample.

\begin{figure*}
	\centering
	\includegraphics[width=\linewidth] {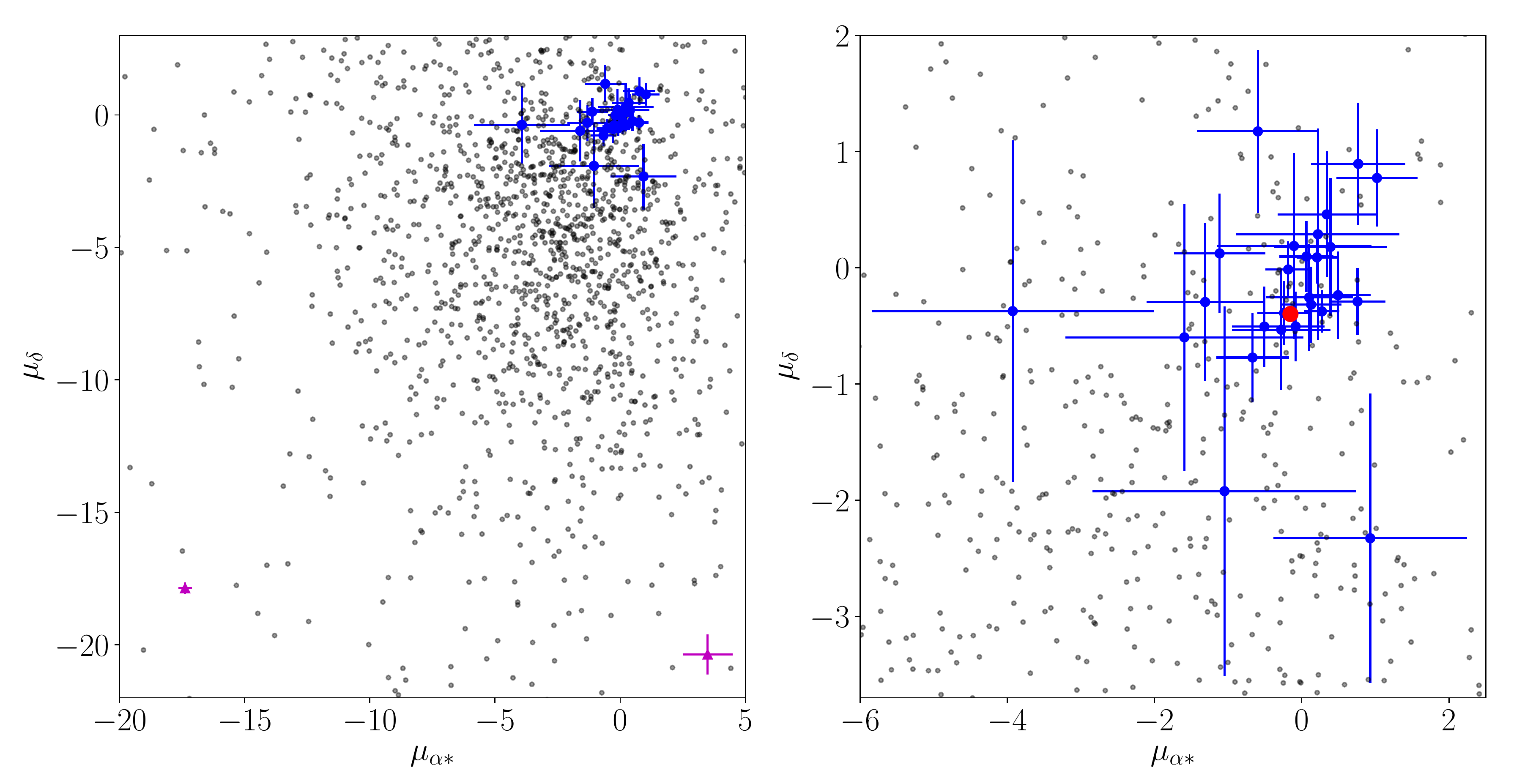}
	\caption{Proper motions for the identified Hercules members, including sources identified in \citet{simon07} and \citet{aden09a}. Left: Proper motions of all 28 sources with a match in \textit{Gaia} DR2, with accepted measurements in blue and the two inconsistent measurements shown as magenta triangles. Right: Close up of the accepted proper motions. The red dot marks the mean proper motion. Proper motions of other \textit{Gaia} sources within 0.3 deg of the centre of Hercules are also shown.}
	\label{fig:gaiapm}
\end{figure*}

\citet{kirby13} define a universal [Fe/H]--luminosity relation for Local Group dwarfs. We show this on a plot of metallicity vs. luminosity for a selection of Milky Way and M31 dwarfs in Fig. \ref{fig:fehlumplot}. The low metallicity of Hercules places it $\sim$1--$\sigma$ below this relation. A dwarf galaxy of Hercules's luminosity would be expected to have a metallicity of $\sim-2.2$ dex. One might expect a disrupting dwarf to be more metal--rich than the relation predicts, as the higher initial mass increases its ability to self--enrich. The metal--poor nature of Hercules is therefore surprising. Other potentially disrupting dwarfs have been found to be fully compatible with this relation (e.g. \citealt{collins17, longeard18}). There does appear to be some scatter about the relation, particularly at the low luminosity end, and the large spread in metallicities may push Hercules closer to the relation.

We might also expect to find a metallicity gradient in our data. To date, most dwarf galaxies have displayed a negative gradient; that is, more metal--rich stars are located at the centre of the galaxy, becoming increasingly metal--poor with distance (e.g. \citealt{battaglia06,koleva09,kirby12,taibi18}). The most probable explanation is that over time, the galaxy becomes increasingly metal--rich due to enrichment from supernovae feedback, whilst gas is preferentially lost from the outer regions, resulting in an increasingly metal--rich and centrally concentrated young population. We find no evidence for a metallicity gradient across Hercules. Our measured gradient of $\frac{d\rm{[Fe/H]}}{d\rm{r}}=-0.01\pm0.06$ dex arcmin$^{-1}$ is consistent with zero within 1$\sigma$.

%Fig. \ref{fig:fehgradient} shows the measured metallicities of our sample as a function of distance from the centre of Hercules. We detect a slight error--weighted metallicity gradient of $\frac{d\rm{[Fe/H]}}{d\rm{r}}=-0.01\pm0.06$ dex arcmin$^{-1}$ across our entire sample. However, we note that this appears to be skewed by the high metallicity, most distant member. If we exclude this star, the error--weighted metallicity gradient becomes $\frac{d\rm{[Fe/H]}}{d\rm{r}}=-0.05\pm0.04$ dex arcmin$^{-1}$. This supports the presence of a mild negative gradient across the galaxy, albeit with large uncertainties due to the small sample size. The excluded star--- object 35 in field H3--- is our most distant identified member of Hercules, located outside the half--light radius. It may be possible that this star is a member of Hercules which has been tidally stripped from the more central regions.

\section{Hercules in Gaia DR2}\label{sec:gaia}

For Hercules to be tidally disrupting as the result of a Milky Way interaction, it must have experienced a recent close passage. Given the large distance between the two galaxies at the present time, this requires Hercules to be on a highly radial orbit \citep{martin10}. We aim to constrain the motion of Hercules using proper motion data from the second \textit{Gaia} \citep{gaia16} data release (DR2, \citealt{gaia18}).

\textit{Gaia} DR2 has already provided strong constraints on the proper motions of many Milky Way satellites, including Hercules (e.g. \citealt{fritz18,kallivayalil18,simon18}). \citet{fritz18} identified 4 member stars from the SG07 and A09 datasets in \textit{Gaia} DR2, from which they determine a proper motion for Hercules of $\mu_{\alpha}^*=-0.297\pm0.118$ mas yr$^{-1}$, $\mu_{\delta}=-0.329\pm0.094$ mas yr$^{-1}$, suggestive of an elliptical orbit with a pericentre of 14--20 kpc, depending on the Milky Way mass. Further analysis of the SG07 and A09 datasets by \citet{fu19} returned a weighted mean proper motion of $\mu_{\alpha}^*=-0.16\pm0.09$ mas yr$^{-1}$, $\mu_{\delta}=-0.41\pm0.07$ mas yr$^{-1}$, consistent with an orbital pericentre of 47 kpc. The inclusion of additional data points provided by the new spectroscopy presented here should help to further constrain the proper motion of Hercules.

For comparison to the DR2 archive, we combine our membership sample with those of SG07 and A09, to create a comprehensive set of 49 Hercules members. We extract all sources within 0.2 degrees of the centre of Hercules from the \textit{Gaia} DR2 archive, to cover the full field of view of our observations. Catalogue matching is used to identify the closest DR2 source (on the sky) to each kinematic member. We select only those with a match within 1 arcsec and with kinematic information available in DR2. As a check, we compare the magnitudes of the kinematic members with the corresponding \textit{Gaia} magnitudes, and confirm a positive correlation between the two. This process returns a sample of 28 sources. We reject two of these sources as their proper motions are more than 3-$\sigma$ outside the mean (see Fig. \ref{fig:gaiapm}, left panel). One of these rejected proper motions is defined as a member in SG07; the other is defined as a member in this work on the basis of its velocity and CMD position (object 35 in field H3). We choose to retain it for our spectroscopic analysis as its inclusion does not significantly affect the results, but note that the proper motion may imply that this star is not a true Hercules member.

\begin{figure}
	\centering
	\includegraphics[width=\linewidth] {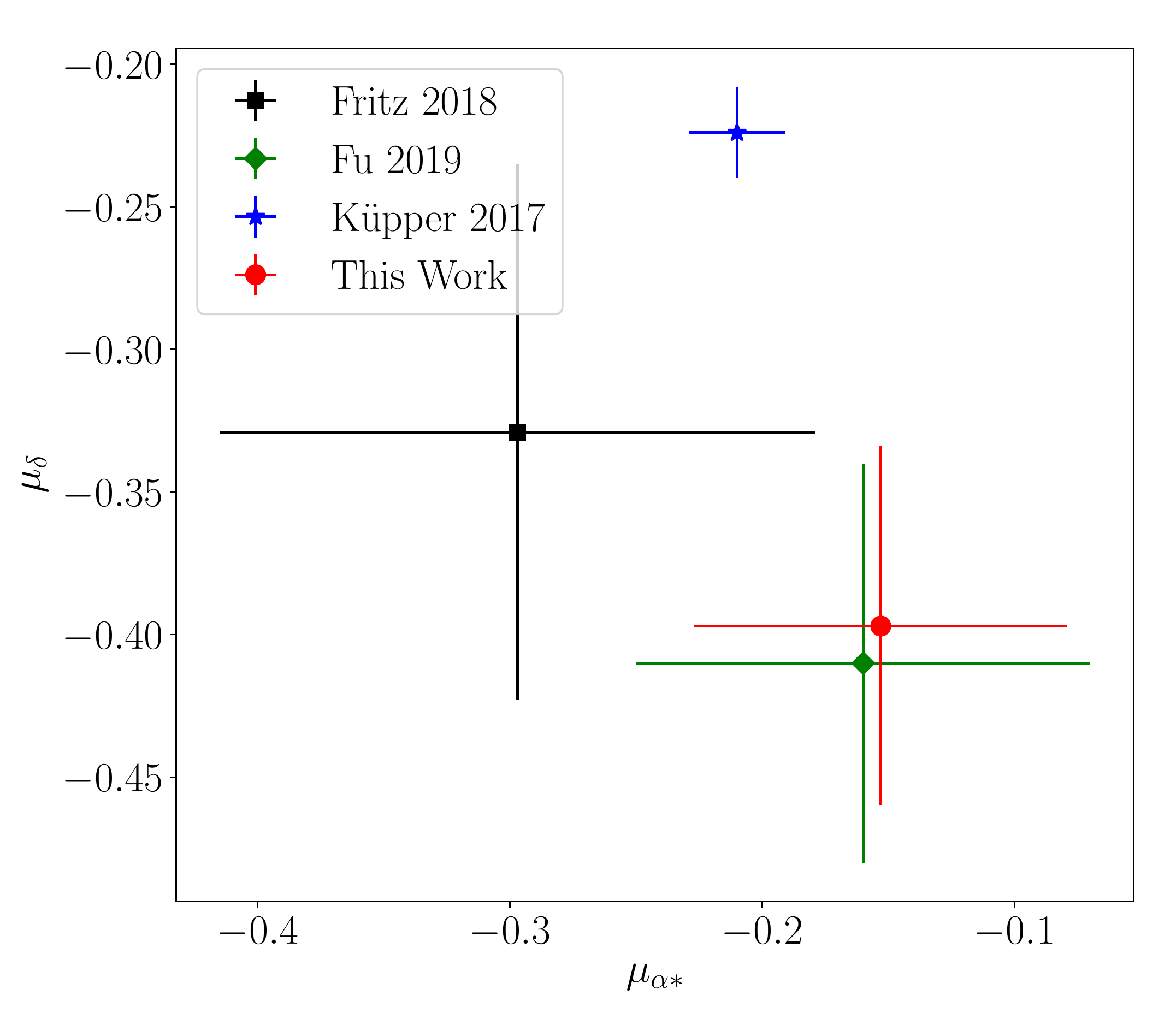}
	\caption{Comparison of the uncertainty--weighted mean proper motion derived in this work with those of \citet{fritz18} and \citet{fu19}. The theoretical proper motion for the predicted orbit of K17 is also shown.}
	\label{fig:meanpm}
\end{figure}

\begin{figure*}
	\centering
	\includegraphics[width=0.9\linewidth] {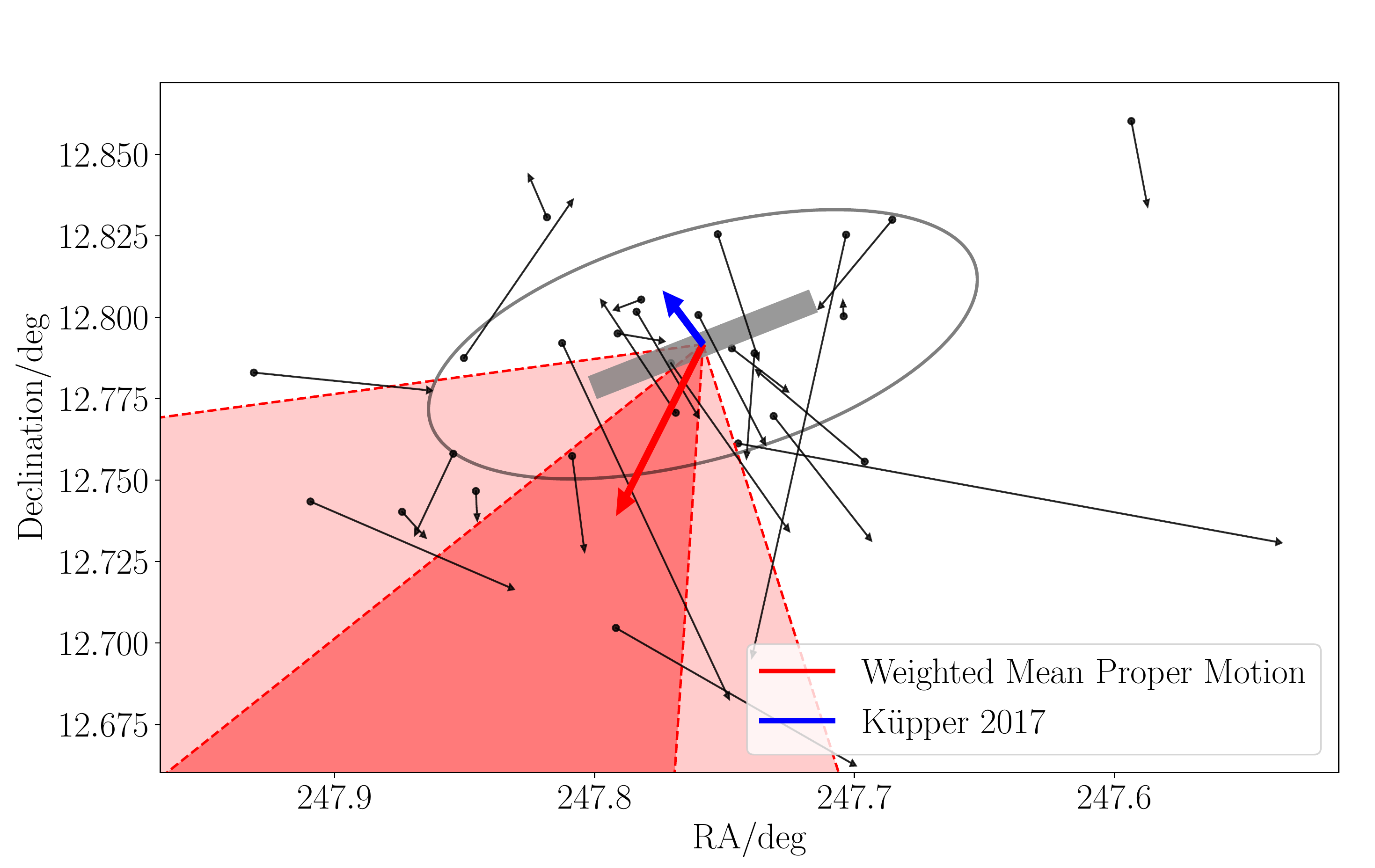}
	\caption{Proper motions for the identified Hercules members, plotted by position in the RA--Dec. plane. The Hercules members with a proper motion measurement in \textit{Gaia} are shown as black arrows. The ellipse marks the half--light radius of Hercules, with the elongation measured in \citet{martin08}; the grey bar indicates the steeper position angle measured by \citet{sand09}. The red arrow marks the weighted mean proper motion of Hercules, with 2--$\sigma$ shaded bands. This proper motion is inconsistent with the prediction of \citet{kupper17} (blue arrow) and with the position angle of Hercules.}
	\label{fig:propermotionmap}
\end{figure*}

To account for the correlation between the proper motions in RA and declination, we fit them simultaneously using a multidimensional Gaussian. We determine the mean proper motion of the galaxy by using \textsc{emcee} to find the proper motions which maximise the likelihood function

\begin{eqnarray}
\mathcal{L}\propto \exp{\bigg(-\frac{1}{2}(\hat{x}-\hat{\mu})^{T}\Sigma^{-1}(\hat{x}-\hat{\mu})\bigg)}
\label{eq:multidimgauss}    
\end{eqnarray}

\noindent where

\begin{eqnarray}
\hat{x} = 
 \begin{pmatrix}
  \mu_{\alpha,i} & 0 \\
  0 & \mu_{\delta,i}
 \end{pmatrix} \hspace{10pt} \rm{and} \hspace{10pt}
\hat{\mu} = 
 \begin{pmatrix}
  \mu_{\alpha} & 0 \\
  0 & \mu_{\delta}
 \end{pmatrix}
\end{eqnarray}

\noindent are vectors describing the individual proper motion of a given source and the mean proper motion of the system respectively. $\Sigma$ is the covariance matrix which accounts for the errors in the measured proper motions ($\delta_{\mu\alpha}$, $\delta_{\mu\alpha}$), and the correlation C$_{\mu\alpha\mu\delta}$ between the two dimensions, and is defined by

\begin{eqnarray}
\Sigma=
  \begin{bmatrix}
  \delta_{\mu\alpha,i} & 0 \\
  0 & \delta_{\mu\delta,i}
 \end{bmatrix} \cdot
 \begin{bmatrix}
  C_{\mu\alpha\mu\delta,i} & 1 \\
  1 & C_{\mu\alpha\mu\delta,i}
 \end{bmatrix} \cdot
 \begin{bmatrix}
  \delta_{\mu\alpha,i} & 0 \\
  0 & \delta_{\mu\delta,i}
 \end{bmatrix}
\label{eq:multidimsigma}    
\end{eqnarray}

\smallskip
From our sample of 26 sources, we determine an uncertainty--weighted mean proper motion of $\mu_{\alpha}^*=\mu_{\alpha}\cos(\delta)=-0.153\pm{0.074}$ mas yr$^{-1}$, $\mu_{\delta}=-0.397\pm0.063$ mas yr$^{-1}$, with a correlation coefficient between them of 0.104. The quoted error on the mean includes the systematic error of 0.035 mas yr$^{-1}$ on proper motion measurements of dSphs identified by \citet{gaia18b} added in quadrature. The individual proper motions are plotted in Fig. \ref{fig:gaiapm}, with the weighted mean proper motion shown as the red dot. We note that if we use only the 9 Hercules members identified in this study which have \textit{Gaia} proper motions (see table \ref{tab:members}), we obtain a mean proper motion of $\mu_{\alpha}^*=\mu_{\alpha}\cos(\delta)=-0.087\pm{0.128}$mas yr$^{-1}$, $\mu_{\delta}=-0.383\pm0.103$ mas yr$^{-1}$, fully consistent with that of the full sample.

In Fig. \ref{fig:meanpm}, we compare our proper motion results to those already found in the literature. Our result is fully consistent with the result of \citet{fu19}, but with reduced uncertainties in accordance with the increased sample size. We also show the proper motion predicted by K17 to reproduce their theoretical orbit. For the extremely radial orbit of K17, a precise proper motion of $\mu_{\alpha}^*=\mu_{\alpha}\cos(\delta)=0.210^{+0.019}_{-0.013}$mas yr$^{-1}$, $\mu_{\delta}=-0.224^{+0.016}_{-0.015}$ mas yr$^{-1}$ is predicted. Within the error ranges, this is consistent with our measurement in the $\mu_{\alpha}$ dimension. However, it is more than 2--$\sigma$ offset in the $\mu_{\delta}$ dimension, suggesting that the K17 model is incompatible with \textit{Gaia} DR2 observations.

Fig. \ref{fig:propermotionmap} illustrates the proper motion of Hercules in the RA--Dec. plane. The black arrows indicate the individual proper motions, corrected to account for the reflex motion of the Sun assuming all sources are located at 132 kpc. The ellipse marks the half--light radius of Hercules with the position angle measured by \citet{martin08}; the grey bar indicates the position angle measured by \citet{sand09}. If Hercules is being tidally disrupted, its elongation should be aligned along the orbital path. However, the bulk proper motion (red arrow) appears inconsistent with the orientation of Hercules by more than 1--$\sigma$, as shown by the red shaded regions. It is also inconsistent with the proper motion required for the `exploding satellite' scenario proposed by K17.

\subsection{Modelling the Tidal Disruption of Hercules} \label{sec:orbit}

In order to test whether the proper motion and radial velocities measured in this work are consistent with the orientation of Hercules, we simulate the tidal disruption of Hercules. This is done using the modified Lagrange Cloud stripping (mLCs) technique from \citet{gibbons14}, which was designed to rapidly reproduce the tidal streams formed during tidal disruption. This method was updated to include the effect of the Large Magellanic Cloud (LMC) in \cite{erkal19a}. This is crucial since recent works (e.g. \citealt{kallivayalil13,penarrubia16,erkal19a}) have found that the LMC has a mass of $\sim 1-2.5\times10^{11} M_\odot$. Such a massive LMC can significantly change the orbit of a number of Milky Way satellites \citep{erkal19}, as well as deflect streams in the Milky Way \citep{erkal18,erkal19a,shipp19}.

For the potential, we used the realistic Milky Way potential from \citet{mcmillan17}, implemented in \textsc{galpot} \citep{dehnen98}, and a $1.5\times10^{11} M_\odot$ LMC modelled as a Hernquist sphere with a scale radius of $17.13$ kpc (as in \citealt{erkal19}). Since the mLCs method includes both the potential of the Hercules progenitor and the LMC, we only used \textsc{galpot} to evaluate the forces from the Milky Way potential and not to integrate orbits. The integration itself was done using a leapfrog method as described in \citet{erkal19a}. We model Hercules as a Plummer sphere with a mass of $2\times10^{5} M_\odot$ and a scale radius of 500 pc. For the LMC, we use the radial velocity from \citet{vandermarel02}, the distance from \citet{pietrzynski13}, and proper motions from \citet{kallivayalil13}. We sample the present day observables (i.e. distance, radial velocities, and proper motions) of both Hercules and the LMC. We rewind Hercules for 5 Gyr in the combined presence of the Milky Way and the LMC. The properties of this orbit (i.e. pericentre, apocentre, and eccentricity) are listed in Table \ref{tab:herc_properties}. This orbit is similar to the results of \citet{fu19}, who measured a slightly different set of proper motions, and who neglected the influence of the LMC. We have checked that if we neglect the LMC, our results are in better agreement with the orbital properties in \citet{fu19}.

\begin{figure}
	\centering
	\includegraphics[width=\linewidth,trim={0.0cm 1.7cm 1cm 2.7cm},clip] {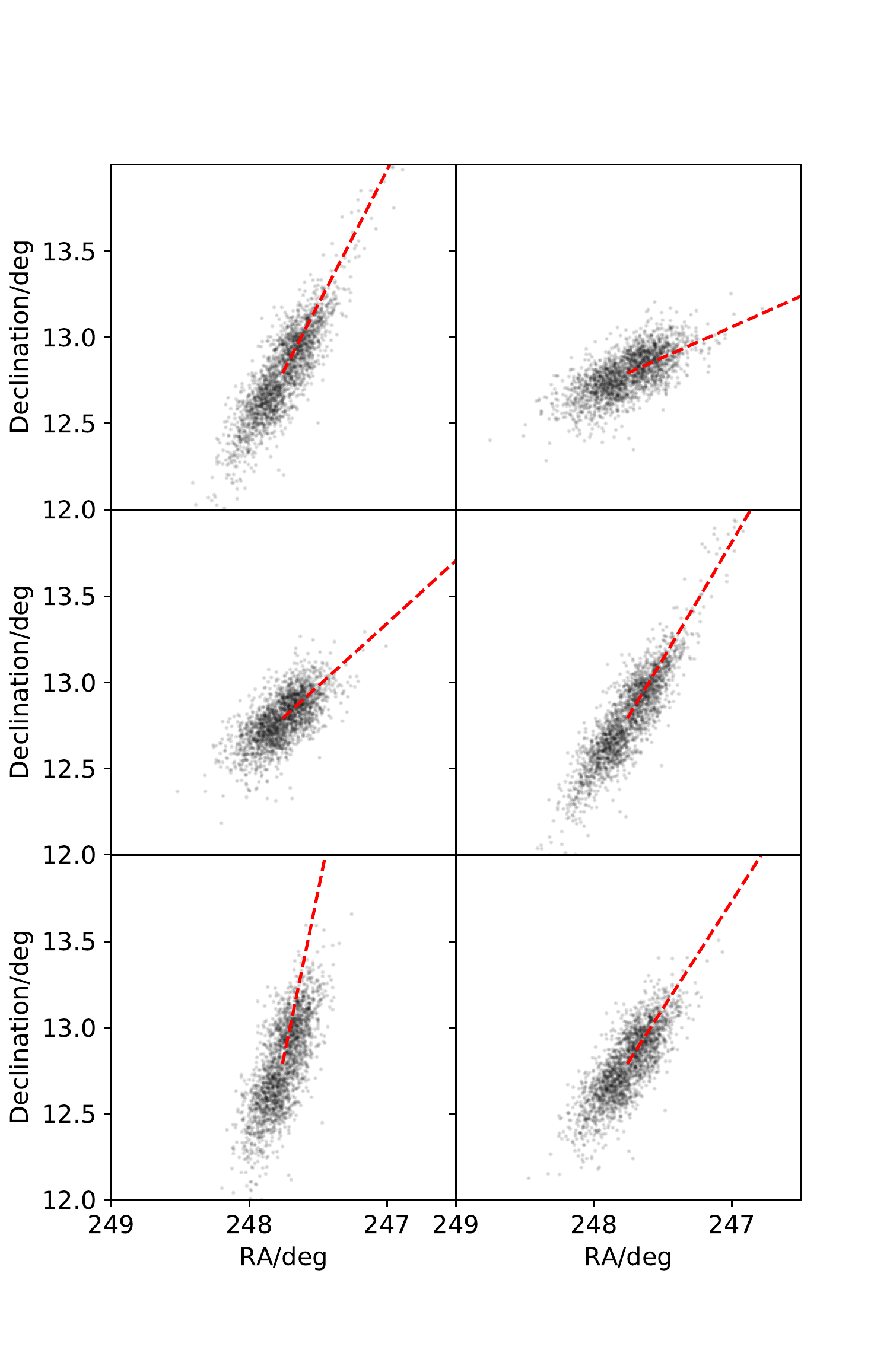}
	\caption{Example of six simulations of the disruption of Hercules. The dashed-red line shows the past orbit of Hercules and the black points show the simulated disruption of Hercules. Interestingly, the elongation of Hercules is always well aligned with the past orbit of Hercules. This is in contrast to observations of Hercules which show the proper motions and orientation are misaligned (see Fig. \ref{fig:propermotionmap}).}
	\label{fig:sim}
\end{figure}

After this rewinding procedure, the Hercules progenitor is then evolved back to the present while disrupting. Fig. \ref{fig:sim} shows 6 such realisations of the disruption of Hercules. The black points show the tidal debris of Hercules and the dashed red line shows the recent orbit of Hercules. By definition, this orbit is aligned with the reflex corrected proper motions of Hercules and thus this figure shows that proper motions are expected to be very well aligned with the tidal debris of Hercules. This is in contrast to Fig. \ref{fig:propermotionmap} which shows that the orientation of Hercules is misaligned with its proper motions.

In order to test how generic this is, we also disrupted Hercules in the \textsc{MWPotential2014} from \citet{bovy15} using the same mLCs technique described above. Although the orbits are slightly different, i.e. larger pericentre and apocentre due to the lower mass of the potential, we find the same generic alignment as in the potential from \citet{mcmillan17}. Finally, since the alignment of Hercules is measured within its half-light radius, we also tested the alignment of Hercules with $N$-body simulations which account for the internal dynamics of Hercules. These simulations were run with the $N$-body part of \textsc{gadget-3}, which is similar to \textsc{gadget-2} \citep{springel05}. We used the \textsc{MWPotential2014} from \citet{bovy15} for the Milky Way but ignored the LMC. Hercules is modelled as a Plummer sphere with a mass of range of either $10^6 M_\odot$ or $10^7 M_\odot$ and a scale radius of 1 kpc, using $10^5$ particles and a softening of 85 pc. We find that the inner region of Hercules is spherical but the outer parts were aligned with the past orbit, as with the debris in Fig. \ref{fig:sim}. Thus, in all cases, the debris is expected to be closely aligned with the proper motions.

There are several plausible explanations for the misalignment between the observed proper motions and shape of Hercules (see Fig. \ref{fig:propermotionmap}). First, the orientation of Hercules is a $\sim$2--$\sigma$ outlier with respect to the proper motions so perhaps improved proper motions will be more consistent with the orientation of Hercules. In order to give the formal tension, we Monte Carlo sample the orientation and proper motion and find a 1.75--$\sigma$ tension with the position angle from \citet{martin08}. Along these lines, we note that other measurements of the shape of Hercules are in better agreement with the proper motions. The position angle determined by \citealt{sand09}, shown as the grey bar in Fig. \ref{fig:propermotionmap}, and the measurement of -67$\pm$0.3$^{\circ}$ by \citealt{roderick15} both reduce the tension, though they are still misaligned by more than 1--$\sigma$. Future observations of the proper motions, e.g. with \textit{Gaia} DR3, will allow us to determine whether they are aligned with the shape of Hercules.

\begin{figure}
	\centering
	\includegraphics[width=\linewidth] {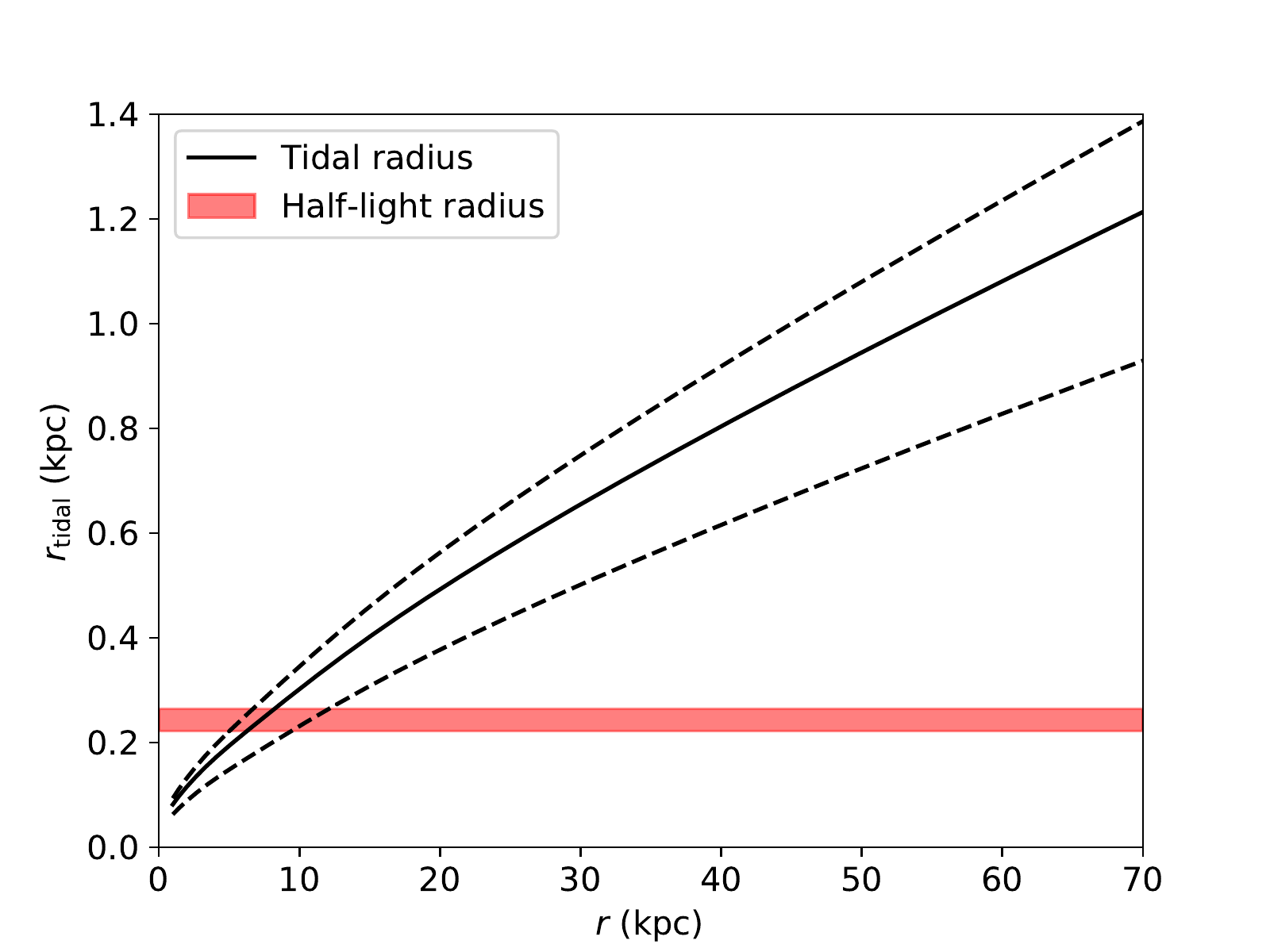}
	\caption{Tidal radius compared to half--light radius as a function of pericentric distance. The tidal radius is computed using the mass (and errors) inferred for Hercules in Sec. \protect\ref{sysvel}. The half-light radius comes from \protect\citet{martin08}. When the half-light radius is similar to the tidal radius, the system should be significantly distorted by tides. Given the large uncertainty on the pericentre of Hercules (see Tab. \protect\ref{tab:herc_properties}) it is unclear from the current data whether Hercules has been tidally distorted by the Milky Way. }
	\label{fig:tidal}
\end{figure}

Based on the mass estimate for Hercules provided in \S\ref{sysvel}, we can also estimate the tidal radius of Hercules in the potential from \citet{mcmillan17}. A comparison between the tidal radius (as a function of pericentric distance) and the half-light radius is shown in Fig. \ref{fig:tidal}. This shows that if the pericentre of Hercules is less than $\sim 12$ kpc, Hercules could be significantly tidally distorted. Although there is a significant uncertainty on the pericentre ($50.9^{+24.2}_{-23.6}$ kpc, see Tab. \ref{tab:herc_properties}), only $\sim2\%$ of the orbits we considered have a pericentre less than 12 kpc. Thus, it appears unlikely that Hercules has been strongly affected by the tides of the Milky Way. 

Alternatively, the shape of Hercules could reflect the fact that it was not originally spherical. Observations have shown that some field dwarfs can display aspherical morphologies (for example WLM; Sag DIG; NGC 3109 \citealt{mcconnachie12}). However, these systems are significantly more massive than Hercules, and as observations of isolated UFDs are limited, it is unclear whether a significant elongation is expected in this mass range.

\section{Conclusions} \label{sec:conc}
 
We have presented new Keck II--DEIMOS spectroscopy of the Hercules dwarf galaxy, and analysed the resulting chemo--kinematics. We measure the heliocentric systemic velocity and velocity dispersion of Hercules to be $v_{\mathrm{Herc}}=46.4\pm1.3$ kms$^{-1}$ and  $\sigma_{v,\mathrm{Herc}}=4.4^{+1.4}_{-1.2}$ kms$^{-1}$ respectively. These are in close agreement with previous measurements. The metallicity of Hercules is measured to be  [Fe/H]= $-2.48\pm0.19$ dex, with a dispersion of $\sigma_{\rm{[Fe/H]}}= 0.63^{+0.18}_{-0.13}$ dex. Hercules is a particularly metal--poor galaxy, falling $\sim$1--$\sigma$ below the standard mass--metallicity relation. This could be a result of scatter about the relation at the low luminosity end, or it may indicate the stronger tidal disruption of lower luminosity dwarfs.

Through comparison to the \textit{Gaia} DR2 archive we also measure the proper motion of Hercules, and find an uncertainty--weighted proper motion of $\mu_{\alpha}^*=\mu_{\alpha}\cos(\delta)=-0.153\pm{0.074}$mas yr$^{-1}$, $\mu_{\delta}=-0.397\pm0.063$ mas yr$^{-1}$, with a correlation coefficient between them of 0.104. Using these proper motions and the observed radial velocity, we model the disruption of Hercules and find that the debris is always well aligned with the present-day proper motions. This is in contrast to the observed proper motions which are slightly misaligned with the position angle of Hercules. We argue that since the misalignment is not very significant, future observations with \textit{Gaia} DR3 may resolve this tension and find a better alignment. Alternatively, it could be that Hercules was originally non-spherical, given that some field dwarfs have aspherical shapes.

\section*{Acknowledgements}
The authors would like to thank Joshua Simon and Marla Geha for providing the \citet{simon07} catalogue. We thank Justin Read and Michele De Leo for helpful discussions. Research by DJS is supported by NSF grants AST-1821967, 1821987, 1813708, 1813466, and 1908972. JS acknowledges support from the Packard Foundation.

The LBT is an international collaboration among institutions in the United States, Italy and Germany. LBT Corporation partners are: The University of Arizona on behalf of the Arizona Board of Regents; Istituto Nazionale di Astrofisica, Italy; LBT Beteiligungsgesellschaft, Germany, representing the Max-Planck Society, The Leibniz Institute for Astrophysics Potsdam, and Heidelberg University; The Ohio State University, and The Research Corporation, on behalf of The University of Notre Dame, University of Minnesota and University of Virginia.

Some of the data presented herein were obtained at the W. M. Keck Observatory, which is operated as a scientific partnership among the California Institute of Technology, the University of California and the National Aeronautics and Space Administration. The Observatory was made possible by the generous financial support of the W. M. Keck Foundation. The authors wish to recognise and acknowledge the very significant cultural role and reverence that the summit of Mauna Kea has always had within the indigenous Hawaiian community. We are most fortunate to have the opportunity to conduct observations from this mountain.

This work has made use of data from the European Space Agency (ESA) mission {\it Gaia} (\url{https://www.cosmos.esa.int/gaia}), processed by the {\it Gaia} Data Processing and Analysis Consortium (DPAC, \url{https://www.cosmos.esa.int/web/gaia/dpac/consortium}). Funding for the DPAC has been provided by national institutions, in particular the institutions participating in the {\it Gaia} Multilateral Agreement. This work makes use of \textsc{emcee}.

%%%%%%%%%%%%%%%%%%%%%%%%%%%%%%%%%%%%%%%%%%%%%%%%%%



\bibliographystyle{mnras}
\bibliography{hercules} % if your bibtex file is called example.bib

%%%%%%%%%%%%%%%%%%%%%%%%%%%%%%%%%%%%%%%%%%%%%%%%%%

%%%%%%%%%%%%%%%%% APPENDICES %%%%%%%%%%%%%%%%%%%%%

\appendix

\section{Spectra of Identified Members}\label{appendixa}

\begin{figure*}
	\centering
	\includegraphics[width=0.95\linewidth] {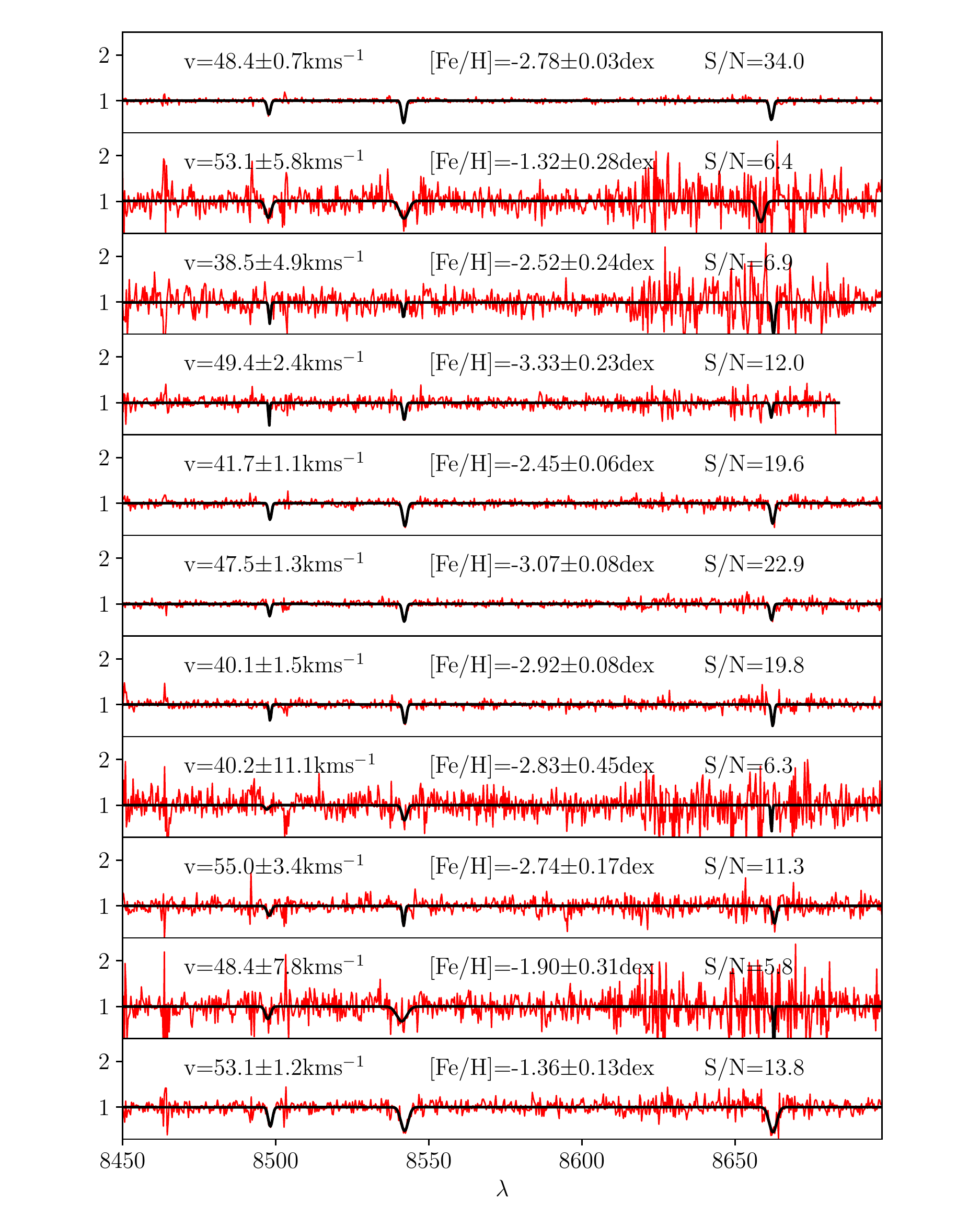}
	\caption{Spectra of the identified Hercules members for which metallicities have been obtained. The black line displays the best Gaussian fit to the Ca II triplet. The measured velocity and metallicity of each object are labelled, as is the S/N of the spectrum.}
	\label{fig:gaussfit}
\end{figure*}

Fig. \ref{fig:gaussfit} shows the normalised spectra of our identified members for which metallicities have been obtained. A black line shows the Gaussian fit to the Ca II triplet. The velocity, metallicity, and S/N of each star are provided. Objects which were too noisy or incomplete in the CaT region to obtain a metallicity measurement are not shown.

\renewcommand{\thefigure}{A1 Cont}
\setcounter{figure}{0}

\begin{figure*}
	\centering
	\includegraphics[width=0.8\linewidth] {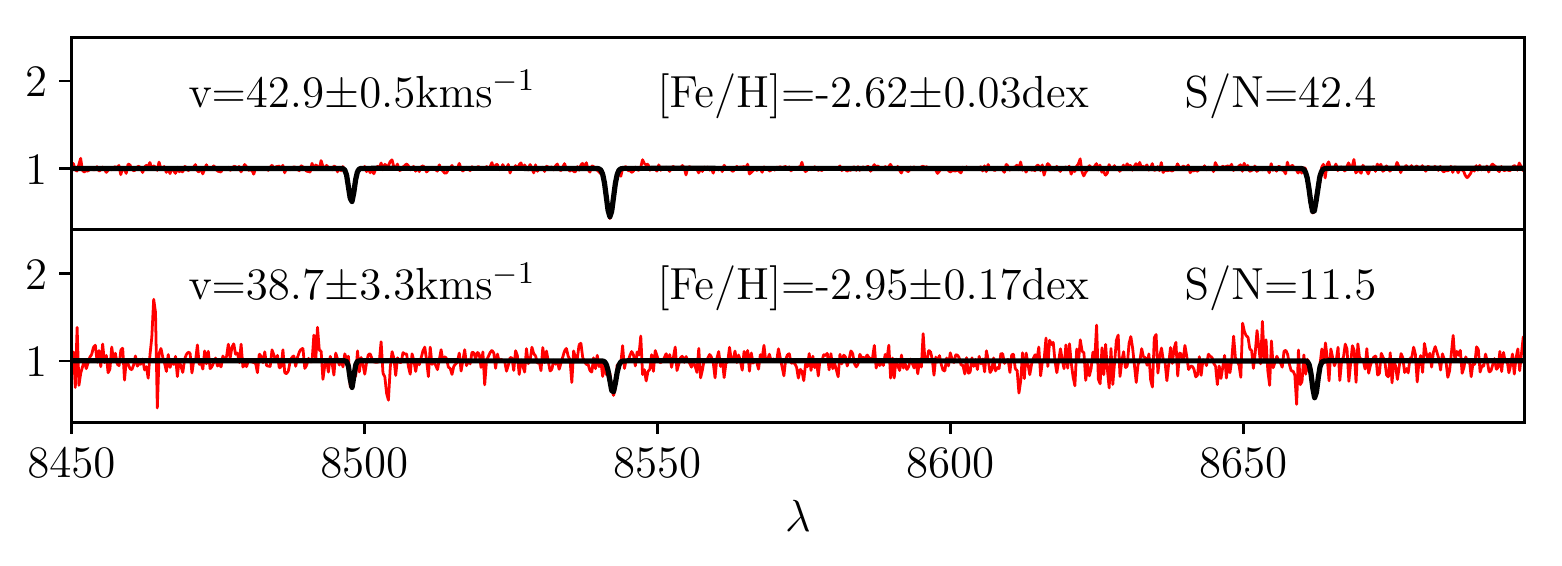}
	\caption{Spectra of the identified Hercules members for which metallicities have been obtained. The black line displays the best Gaussian fit to the Ca II triplet. The measured velocity and metallicity of each object are labelled, as is the S/N of the spectrum.}
	\label{fig:gaussfit2}
\end{figure*}

%%%%%%%%%%%%%%%%%%%%%%%%%%%%%%%%%%%%%%%%%%%%%%%%%%

% Don't change these lines
\bsp	% typesetting comment
\label{lastpage}
\end{document}